\documentclass[10pt]{article}
\usepackage{manuscript}
\hypersetup{pdftitle={When Does Evidence-Graph Structure Predict Answer Correctness? Label-Free Abstention for Structured Retrieval},pdfauthor={Anjan Goswami}}
\title{When Does Evidence-Graph Structure Predict Answer Correctness? Label-Free Abstention for Structured Retrieval}
\author{Anjan Goswami\thanks{Email: \href{mailto:anjan@smartinfer.com}{\texttt{anjan@smartinfer.com}}. ORCID: \href{https://orcid.org/0009-0002-1778-3454}{0009-0002-1778-3454}.}\\SmartInfer, Inc.}
\date{}
\begin{document}
\maketitle
\begin{abstract}
Retrieval systems return scored answers, but a high score does not show that an answer should be trusted. The proposed training-free abstention method applies the boundary cut to the selected evidence in an evidence graph. The signal is the weighted cut between selected and unselected candidates and requires no correctness labels or confidence-model training. On MemoryAgentBench (MAB) FactConsolidation multi-hop, with unresolved queries ranked last for every selector, it raises area under the precision--coverage curve from $0.736$ for the best score baseline to $0.785$ ($+0.049$, 95\% CI $[+0.008, +0.092]$); under the native evaluation, where score baselines keep their unresolved values, the gap is $+0.108$. A zero-cut subset contains $71$ of $400$ queries at $100\%$ observed precision. In this pipeline, the subset corresponds to structural closure; controlled analyses attribute much of the remaining separation to resolution status, with little benefit from continuous edge weights. Across external benchmarks, usefulness varies with graph construction, edge semantics, and relation; typed structure alone does not ensure useful confidence. A supervised gradient-boosted selector achieves $0.814$ versus the boundary cut's $0.785$ when the retrieval stack matches. These findings provide empirical guidance for using available evidence graphs for lightweight, auditable post-retrieval abstention without additional model inference.
\end{abstract}

\section{Introduction}
\label{sec:intro}

Retrieval systems often decide whether to trust an answer from the
retrieval score alone. But a high score does not tell us whether the
selected evidence is uncontested.

Consider the query: \emph{What country is Alice's birthplace in?}
A retrieval system may select the facts ``Alice was born in Paris'' and
``Paris is in France'' and return \emph{France}. A score-based confidence
rule asks whether these selected facts received high scores. It does not
ask whether the same local evidence graph also contains a competing
birthplace, a conflicting path, or other candidates connected to the
selected support.

This distinction matters in structured retrieval systems. Agent-memory,
knowledge-graph, and provenance-aware pipelines often already organize
retrieved evidence as typed graphs. These graphs contain information that
a scalar retrieval score discards: how the selected evidence is connected
to plausible alternatives.

We ask whether this structure can be used directly to decide when a
retrieved answer should be trusted.

Existing query performance prediction methods estimate retrieval quality from scores or relationships among ranked items (\S\ref{sec:related}). We instead study whether the boundary of the evidence actually selected to support an answer provides a query-level signal for abstention in evidence graphs.

Our starting point is simple. Let $P$ be the evidence selected for the returned answer, and let the local evidence graph contain both $P$ and nearby unselected candidates. The boundary measures how strongly the selected evidence remains connected to unselected candidates in the evidence graph. A larger boundary indicates that the selected support is less structurally isolated; whether this separation predicts correctness depends on the semantics of the graph edges.

The resulting abstention rule can be added to a structured retrieval pipeline without training a confidence model. It leaves the retrieval and answer-selection process unchanged, then uses a sparse edge sum over the available local evidence graph to decide whether to return the answer or abstain. This separates answer selection from answer acceptance, and offers lightweight post-retrieval abstention without additional model inference.

The central question is:

\begin{quote}
\emph{When does the structure of retrieved evidence provide information
about whether the returned answer is correct?}
\end{quote}

Our experiments show that the answer depends on what the graph represents.
The signal is useful when edges encode \emph{support compatibility}: two
connected candidates are expected to support the same answer hypothesis.
Typed entity--relation structure can provide this property, whereas
temporal adjacency, co-occurrence, or generic semantic similarity often do
not. Our experiments identify alignment with answer-support semantics as a key condition for useful structural confidence.

We study the weighted boundary between the evidence selected for an answer and the remaining candidates. We call this boundary statistic the \emph{boundary cut}; it is the Dirichlet energy of the selected-support indicator (\S\ref{sec:coherence-energy}), and it appears as `coherence' in table labels, figure legends, and released artifacts.

On MAB multi-hop, ranking unresolved queries last for every selector gives the boundary cut AUPCC $0.785$ versus the strongest score baseline's $0.736$, a status-matched advantage of $+0.049$ (95\% CI $[+0.008, +0.092]$). The native gap is $+0.108$, explicitly including the difference in unresolved-status handling. The zero-cut subset contains $71/400$ evaluations, all correct. A matched-coverage score threshold selects a substantially different set and overlaps with only 14 of these queries.

The experiments also reveal why the effect occurs. On MAB, the
100\%-precision regime corresponds to \emph{structural closure}: the
selected support exhausts its local typed neighborhood. Structural controls match the weighted boundary signal on MAB, but its
apparent advantage outside closure largely disappears after unresolved
queries are excluded. External benchmarks then show the limit of this result: the same
boundary statistic is not universally useful, and its behavior changes
with graph construction, candidate-space structure, and edge semantics.

These results identify deployment tradeoffs between structural abstention and learned confidence. GBT has a higher in-distribution point estimate, but its paired difference from the boundary cut is not statistically resolved. The structural rule is recomputed from the deployed evidence graph and requires neither correctness labels nor confidence-model retraining.

\paragraph{Contributions}
\begin{enumerate}[leftmargin=1.2em,itemsep=2pt]
\item \textbf{Structural abstention.} A training-free acceptance rule on selected-support boundaries includes a structural-closure regime with $100\%$ observed precision at approximately $18\%$ coverage on MAB multi-hop. This regime is stable across the evaluated encoder, scorer, and $\gamma$ settings and degrades gracefully under extraction noise. Closure is produced by the upstream resolver; it is an empirical reliability indicator, not a certificate of correctness.
\item \textbf{Mechanism.} An empirical decomposition separates closure, resolution state, boundary topology, and continuous edge weights. On MAB, the graded signal outside closure largely reflects resolution-state composition, and continuous weights are unnecessary for the observed selective behavior.
\item \textbf{Applicability.} We identify structural characteristics associated with effective abstention and investigate how they vary across retrieval tasks and answer relations. Cross-benchmark and relation-level evidence links usefulness to graph semantics and candidate-space structure; support compatibility is an explanatory hypothesis rather than a validated predictive criterion.
\item \textbf{Deployment tradeoffs.} On the regenerated population, the boundary cut exceeds four local score-distribution QPP adaptations with paired CIs excluding zero. GBT has the higher in-distribution point estimate, not statistically resolved, and portability across retrieval stacks is mixed. The structural rule needs no correctness labels.
\end{enumerate}

\section{Method}
\label{sec:method}

\subsection{Candidates, Scores, and the Evidence Graph}

For a query $q$, a retrieval system returns candidate items
$\mathcal N=\{o_1,\dots,o_n\}$. A candidate may be a sentence, a
knowledge-graph triple, or an entity, depending on the task. Each
candidate $i$ has a representation $\mathbf v_i\in\mathbb R^d$ and a
retrieval score $b_i$. We write the score vector as
$\mathbf b\in\mathbb R^n$.

The abstention rule also uses a weighted graph
$G_{\mathcal N}=(\mathcal N,E_{\mathcal N},w)$ over the retrieved
candidates. Its adjacency matrix is $A=[w_{ij}]$, with $w_{ij}\ge 0$,
and its Laplacian is $L=D-A$, where $D_{ii}=\sum_j w_{ij}$. An edge
does not mean that two candidates are generally similar. It means that,
for the current query, the two candidates should support the same
answer. A boundary edge therefore connects selected support to an unselected candidate expected to support the same answer hypothesis; it does not by itself encode a contradiction or a competing answer. The method does not prescribe a specific encoder, scorer, or
graph-construction implementation, subject to the support-compatibility
requirement below; the abstention rule itself requires only the resolved
selected support $P$ and the weighted graph. Retrieval scores are upstream
inputs, used to resolve $P$ and for the score-based baselines.

\paragraph{Graph construction and support compatibility}
The abstention rule does not prescribe how the graph is constructed, but its
usefulness depends on what the edge weights mean. In an existing
knowledge graph, raw weights may encode similarity, proximity, or
relation strength. For abstention, these weights can be preserved but
gated by a query-dependent support function:
$w^{\mathrm{verify}}_{ij}=s_{ij}\,m(q,o_i,o_j)$,
where $s_{ij}\ge 0$ is the original edge weight and
$m(q,o_i,o_j)\in[0,1]$ measures whether candidates $o_i$ and $o_j$
should receive similar support scores for the current answer. More
generally, the abstention rule can use any nonnegative application-specific
edge function
$w_{ij}=g(q,o_i,o_j,\mathbf v_i,\mathbf v_j)\ge 0$.
We investigate whether assigning edge weight to candidate pairs whose support scores should agree is associated with informative structural confidence.

To make this condition explicit, let $y_i$ be the latent
answer-support class of candidate $i$: $y_i=y_j$ means that candidates
$i$ and $j$ support the same underlying answer hypothesis. For a graph
with nonzero total edge weight, define
\begin{equation*}
    \rho(G_{\mathcal N};y)
    =
    \frac{\sum_{i<j:\,y_i\neq y_j} w_{ij}}
         {\sum_{i<j} w_{ij}}.
\end{equation*}
A graph constructor is more \emph{support-compatible} when this fraction
is smaller: most edge weight lies within answer-support classes rather
than between classes whose scores should differ. The latent classes
$y_i$ are not observed at deployment and are not inputs to the abstention rule.
They describe a structural characteristic for interpreting graph constructors; this ratio is not a validated predictive criterion.

\subsection{Boundary Cut and Abstention}
\label{sec:coherence-energy}

Let $P$ denote the evidence selected for the returned answer and $z_i=\mathbf 1[i\in P]$ its binary indicator on the local neighborhood $V$. Classical Dirichlet energy is then exactly the weighted boundary cut of the selected support:
\begin{equation}
 E(\mathbf z)=\mathbf z^\top L\mathbf z
 =\sum_{i\in P,\,j\notin P}w_{ij}.
 \label{eq:cut}
\end{equation}
Undirected edges are counted once. On MAB, $P$ is the selected support produced by the benchmark resolver: the resolved multi-hop evidence path or the selected single-hop supporting fact with its induced local support. In the other evaluated settings, $P$ is the selected top-$k$ set (\S\ref{sec:setup}). Thus a low cut means the selected support is weakly coupled to unselected
candidates, and exact zero means no positive-weight edge leaves the
selected support. We later find that, on MAB, continuous edge magnitudes are not
necessary for the selective effect: replacing them with an unweighted
boundary count gives equivalent performance (\S\ref{sec:structural-controls};
see \S\ref{sec:cross-benchmark}).

\begin{algorithm}[t]
\caption{Boundary-cut abstention for one query}
\label{alg:coherence-verification}
\begin{algorithmic}[1]
\Require query $q$, candidate set $\mathcal N=\{o_1,\dots,o_n\}$, candidate representations $\{\mathbf v_i\}_{i=1}^n$, retrieval scores $\mathbf b$, graph constructor $\mathcal G$, selection/resolution rule $\mathcal S$, threshold $\tau$
\Ensure \textsc{commit} or \textsc{abstain}
\State $G_{\mathcal N}\gets \mathcal G(q,\mathcal N,\{\mathbf v_i\}_{i=1}^n)$
\State $(P,o_a)\gets \mathcal S(q,\mathcal N,\mathbf b,G_{\mathcal N})$ \Comment{$\mathcal S$ may ignore $G_{\mathcal N}$}
\If{the resolver reports \textsc{not\_found}}
    \State \Return \textsc{abstain}
\EndIf
\State $z_i \gets \mathbf 1[i\in P]$ for $i=1,\dots,n$ \Comment{hard selected-support field}
\State $E \gets \sum_{i\in P,\,j\notin P} w_{ij}$ \Comment{boundary cut of $P$}
\If{$E\le \tau$}
    \State \Return \textsc{commit}$(o_a,E)$
\Else
    \State \Return \textsc{abstain}$(o_a,E)$
\EndIf
\end{algorithmic}
\end{algorithm}

Unresolved queries are abstained from before evaluating the boundary rule.
For precision--coverage evaluation, they receive a worst-confidence sentinel
and are ranked last.

The abstention rule does not change the retrieved answer; it decides whether to
trust the top-scoring answer. The threshold $\tau$ is
swept to produce precision--coverage curves; the zero-cut result
corresponds to $\tau=0$. Because the cut scales with the edge
weights, fixed absolute thresholds are weight-scale dependent; our
evaluation therefore uses threshold sweeps rather than a universal
threshold. Computing the score requires no correctness labels, but choosing
a threshold to meet a target precision requires separate validation. The
zero-cut rule supplies a fixed structural criterion; its relationship to
correctness is empirical. Laplacian-smoothed scores,
$\tilde{\mathbf b}=(I+\gamma L)^{-1}\mathbf b$, are not part of the
abstention rule; they are included only as a score-based baseline
(Appendix~\ref{app:gamma}).

After the selected support and the graph are available, abstention
is a single sparse edge sum. It costs $O(|E_{\mathcal N}|)$ time. If the
edge list is streamed or already stored, the computation uses $O(1)$
working memory beyond the graph and node field; materializing the sparse
Laplacian costs $O(|E_{\mathcal N}|)$ memory. The dense worst case is
$O(n^2)$. Graph construction is separate from the
abstention rule: a typed graph uses available entity--relation structure, while
a semantic $k$NN graph has the cost of the nearest-neighbor routine used
to build it.

The method applies classical Dirichlet energy to the binary selected-support indicator, yielding the weighted boundary statistic in Eq.~\ref{eq:cut}. It measures how strongly selected evidence is connected to unselected support and supplies an abstention signal for the retrieved support neighborhood.

\subsection{Boundary Interpretation}
\label{sec:basic-properties}
\begin{remark}[Binary selected-support boundary]
\label{rem:boundary}
For a binary selected-support indicator and nonnegative edge weights, zero cut means that no positive-weight edge crosses from $P$ to its complement. Equivalently, the indicator is constant on each positive-weight connected component, and $P$ is a union of such components. Increasing a crossing edge's weight increases the cut by the same amount; increasing an internal edge's weight leaves it unchanged. These are immediate consequences of the boundary sum in Eq.~\ref{eq:cut} and do not establish answer correctness.
\end{remark}

Remark~\ref{rem:boundary} explains why graph construction matters.
Increasing weight on a pair increases its contribution in proportion to
the squared field disagreement. For the selected-support indicator
$z_i=\mathbf 1[i\in P]$, $(z_i-z_j)^2=1$ exactly when one endpoint lies
in $P$ and the other does not, so increasing the weight of such an edge
increases the selected-support boundary cut; this is the primary
interpretation of the remark in our experiments. Adding weight is
useful when the pair should agree under the current answer hypothesis,
but harmful when the pair is only temporally adjacent, co-occurring, or
generically similar. A support gate $m(q,o_i,o_j)$ is therefore useful
because it preserves raw graph weights only on pairs that should
support the same answer.

\subsection{Why Graph Structure Matters}
\label{sec:graph-structure}
The preceding properties hold for whatever abstention weights the constructor
produces, so their practical consequence is a requirement on graph construction:
edges should connect candidates whose support scores should agree under the
current answer hypothesis, and large weights on support-incompatible pairs
(candidates that are merely temporally adjacent, co-occurring, or generically
similar) add boundary cut unrelated to correctness. Appendix~\ref{app:population-vs-selected}
gives a continuous latent-class model that makes this requirement precise by
decomposing Dirichlet energy into within-class and across-class terms; it is
motivational intuition only and is \emph{not} instantiated in any reported
experiment (every reported boundary cut result verifies the hard selected-support
indicator of Eq.~\ref{eq:cut}).

\section{Experimental Setup}
\label{sec:setup}

\paragraph{Experimental rationale and baselines}
The experiments test whether the boundary cut is a label-free abstention
signal when the evidence graph is support-compatible. MAB provides the
positive setting. We compare the boundary cut to three kinds of alternatives. Score-based
confidence baselines (raw score, smoothed score, and score margin) test
whether the boundary cut adds information beyond retrieval confidence alone.
Search-style baselines (leaf MCTS, chain-conditioned MCTS, and
path-score reranking) test whether additional post-retrieval computation
solves the same failures. Learned confidence baselines (logistic
regression and a shallow gradient-boosted tree ensemble) test how a
zero-fit graph signal compares with supervised confidence models trained
on retrieval-time features. In addition, graph ablations and retrieval-stack
shifts test the graph and the encoder/scorer dependence (below).

\paragraph{Statistical unit and metric}
Tables and figure legends label the boundary-cut signal `Coherence' for continuity with the released artifacts.
The reconstructed mapping (Appendix~\ref{app:cross-config}) contains 374 distinct exact question texts across
400 evaluations at four context scales; resampling and folds group all occurrences
of each text together. The mapping uses positional lookup into the available
dataset snapshot; the historical run did not record its dataset revision
(Appendix~\ref{app:cross-config}). Selective prediction is summarized
by AUPCC: predictions are ordered most- to least-confident, and at prefix $k$ we
take precision (correct among the top $k$, over $k$) and coverage $k/n$; AUPCC is
the trapezoidal area under this curve. AUPCC is the primary metric because
the boundary cut is an abstention signal---it changes which predictions are trusted,
not the base retrieval answer at full coverage---so we report both retrieval
accuracy and precision--coverage quality, with bootstrap 95\% CIs over queries
for key comparisons. The boundary cut is oriented so that a lower cut is more confident; raw, smoothed, and margin scores use higher-is-confident;
predictions are sorted by \texttt{argsort}$(-\text{confidence})$ and exact ties are
resolved in the evaluator's native row order (no semantic secondary key). Two
implementations are used: the headline Table~\ref{tab:main} evaluator subsamples the
$n$ prefixes to 50 points before integration, while the learned-baseline and
bootstrap comparisons use all $n$ prefixes; on the headline signals the two agree to
within $0.0024$ AUPCC. Continuous signals (boundary cut, raw, smoothed) are effectively
insensitive to tie ordering; discrete or constant structural-control signals (e.g.\
the closure indicator, complement size, or unweighted boundary count on the
$P\neq V$ subset) can be tie-order-sensitive, so their absolute AUPCC values are
interpreted descriptively and compared as paired deltas under a shared ordering.
Unresolved (\textsc{not\_found}) queries take a worst-confidence sentinel for the boundary cut that ranks them
last; the score baselines carry their native small values and are not forced last,
and the few missing score-derived features are zeroed by the historical
implementation.

\paragraph{Primary benchmark}
MemoryAgentBench (MAB) FactConsolidation~\citep{hu2026memoryagentbench}
is the primary positive test because it provides retrieved candidate
facts, explicit multi-hop evidence, and typed entity--relation structure.
The multi-hop regime tests whether the boundary cut helps when evidence must be
composed across facts; the single-hop regime tests whether the signal
remains useful when base retrieval accuracy is already high.
MAB has four context scales (6k--262k tokens), 100 queries
per split, and 400 queries per regime. 
On MAB, we use one such graph constructor: a typed entity--relation
mask multiplied by squared embedding similarity,
$w_{ij}=M_{ij}(\mathbf v_i^\top \mathbf v_j)^2$.
For MAB, $M_{ij}=1$ when the retrieved candidate facts share the same
extracted entity id, and $0$ otherwise. These entity ids are available in
candidate metadata at retrieval time; the mask does not use gold answers,
correctness labels, or calibration data.

\paragraph{Node field}
In every experiment the node field is a binary selected-support
indicator; no experiment computes the cut of a continuous score field. On
MAB, $P$ is the resolver's selected support as defined in \S\ref{sec:coherence-energy};
on the transfer benchmarks (CoDEx-S, MetaQA-3hop, LongMemEval, LoCoMo), $P$ is
the top-$k$ retrieved set under the smoothed ranking.

\paragraph{MAB resolver and neighborhood}
The MAB multi-hop pipeline runs in order: (1) context parsing/triple extraction into a
per-example knowledge base; (2) question decomposition into a hop chain; (3) per-hop
candidate scoring by smoothed similarity; (4) path resolution, where each hop's object
seeds the next hop, yielding the resolved support path $P$ (\textsc{not\_found} if a hop
fails); (5) construction of the local neighborhood $V$ as $P$ together with its typed
one-hop neighbors, so $P\subseteq V$ by construction; (6) typed edges over $V$; (7)
the boundary cut of the selected-support indicator; (8) correctness
scoring against the gold answer. Gold answers are used only in step~8; they do not enter
the construction of $P$ or $V$.

\paragraph{Transfer and graph ablations}
To test whether the boundary cut depends on a particular retrieval stack, we
evaluate two additional multi-hop configurations: a linear scorer
$\mathbf{q}^\top \mathbf{v}_i$ instead of the default
$(\mathbf{q}^\top \mathbf{v}_i)^2$, and a MiniLM encoder
($d{=}384$) instead of the default Qwen encoder ($d{=}128$), with the
typed graph held fixed (Table~\ref{tab:transfer}). To isolate graph construction itself, we cache
the retrieved MAB candidates, node scores, embeddings, and correctness
labels from one run and recompute the boundary cut post hoc under alternative
graphs: the typed entity--relation graph, a semantic kNN graph, and
hybrid graphs that interpolate the two. Because the candidates and
scores are fixed, this ablation isolates the graph effect; its absolute
AUPCC values are not directly comparable to the end-to-end results in
Tables~\ref{tab:main} and~\ref{tab:learned}.

\paragraph{Use of AI tools}
Experiment code, statistical analyses, and reproducibility audits were implemented with LLM-based coding agents (Claude Code and Codex) under the author's direction. All protocols, eligibility rules, and hypotheses were specified by the author before results were computed; all reported numbers were verified by the author against frozen artifacts; artifact identifiers and protocol records are available from the author. The fact-extraction and question-decomposition model inside the evaluated pipeline is gpt-4.1-mini-2025-04-14 (\S\ref{sec:setup}). LLM tools were also used for writing assistance.

\section{Results}
\label{sec:results}

\subsection{The Boundary Cut Improves Aggregate Precision--Coverage on MAB}
\label{sec:results-mab}

Table~\ref{tab:main} summarizes both regimes (per-split detail in Appendix~\ref{app:splits}).

\begin{table}[t]
\centering
\caption{Aggregate results with the status-matched multi-hop comparison primary: unresolved queries are ranked last for every selector. Native comparisons are reported below the table; resolved-only results are in \S\ref{sec:results-mab}.}
\label{tab:main}
\begin{tabular}{lcccc}
\toprule
Regime & Base Accuracy & Coherence AUPCC & Status-matched smoothed & Gap \\
\midrule
Multi-hop  & 0.420--0.700 & 0.785 & 0.736 & +0.049 \\
Single-hop & 0.900--0.960 & 0.975 & \textemdash & \textemdash \\
\bottomrule
\end{tabular}
\par\smallskip
\begin{minipage}{0.98\linewidth}\footnotesize
Native evaluation (score baselines retain their unresolved values): multi-hop smoothed AUPCC $0.677$, gap $+0.108$; single-hop smoothed AUPCC $0.941$, gap $+0.034$. No status-matched single-hop result is available in the frozen tables, so its status-matched entries are marked \textemdash; its boundary cut AUPCC is the unchanged native value.
\end{minipage}
\end{table}

\begin{figure}[tbp]
\centering
\includegraphics[width=5in]{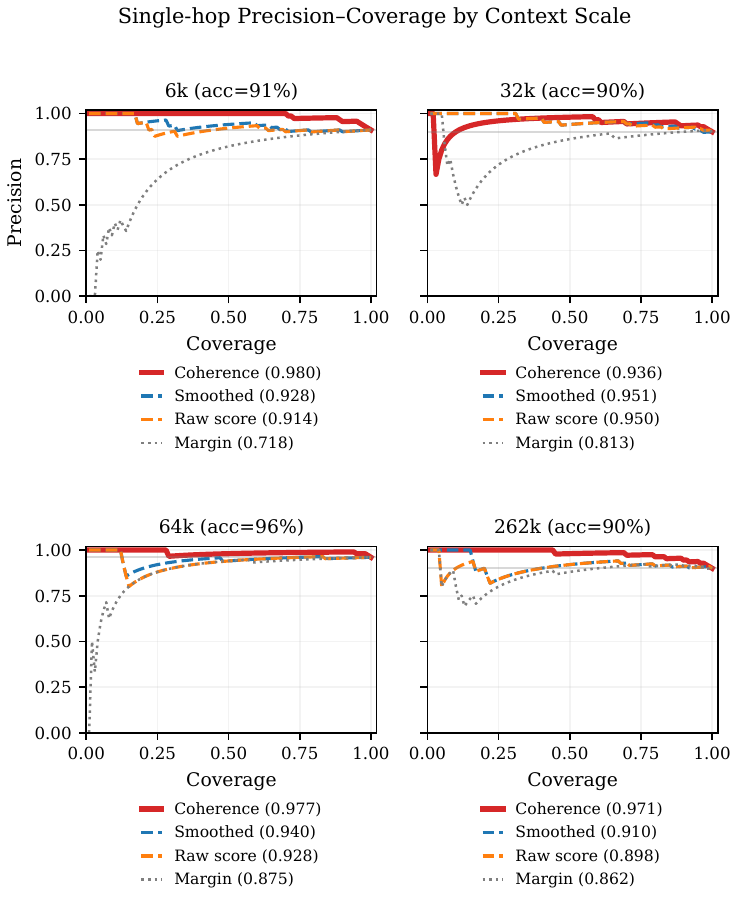}
\Description{Four single-hop panels plot precision against coverage at 6k, 32k, 64k, and 262k context lengths. Each compares the boundary cut, smoothed score, raw score, and margin. The boundary cut has the highest AUPCC at 6k, 64k, and 262k; smoothed score is highest at 32k. All curves end at the base accuracy at full coverage.}
\caption{Single-hop precision--coverage. The boundary cut helps even at 90--96\% base accuracy.}
\label{fig:sh-pc}
\end{figure}

In the multi-hop regime, 71 of 400 rows have a zero boundary cut and all 71 are correct. Under the reconstructed mapping (Appendix~\ref{app:cross-config}), these rows represent 65 distinct questions, all correct. The exact two-sided 95\% Clopper--Pearson interval over distinct questions is $[0.945, 1.000]$. Observed row coverage is 17.75\% (71/400), with 95\% CI grouped by exact question $[0.138, 0.219]$. The single-hop regime shows that the signal is not limited to low-accuracy
retrieval: even with 90--96\% base accuracy, the boundary cut gives the best
aggregate AUPCC. On one split (SH 32k), smoothed score narrowly leads (0.951 vs.\ 0.934); this is the only exception across both regimes. Per-split single-hop precision--coverage curves are shown in Figure~\ref{fig:sh-pc}.

\begin{table}[tbp]
\centering
\caption{Per-split single-hop AUPCC ($n{=}100$ per split). Smoothed score leads on 32k.}
\label{tab:per-split-sh}
\begin{tabular}{lccccc}
\toprule
Split & Accuracy & Coherence & Smoothed & Raw & Margin \\
\midrule
6k   & 0.910 & \textbf{0.980} & 0.928 & 0.914 & 0.718 \\
32k  & 0.900 & 0.936 & \textbf{0.951} & 0.950 & 0.813 \\
64k  & 0.960 & \textbf{0.977} & 0.940 & 0.928 & 0.875 \\
262k & 0.900 & \textbf{0.971} & 0.910 & 0.898 & 0.862 \\
\midrule
All  & 0.917 & \textbf{0.975} & 0.941 & 0.932 & 0.823 \\
\bottomrule
\end{tabular}
\end{table}

This zero-cut subset is not just a repackaging of easy high-score examples. The full overlap analysis (Table~\ref{tab:zero-overlap-app}) compares it with matched-coverage and top-decile score thresholds. A raw-score threshold matched to the zero-cut coverage attains only 67.6\% precision and overlaps with the zero-cut set on just 14 queries (Jaccard 0.11). At matched coverage, the boundary cut selects 57 correct queries that the raw-score
threshold misses.

\begin{table}[tbp]
\centering
\caption{Zero-energy overlap analysis on multi-hop MAB.}
\label{tab:zero-overlap-app}
\begin{tabular}{lccc}
\toprule
Subset & Size & Precision & Overlap w/ zero-energy \\
\midrule
Zero-energy ($E{=}0$) & 71 & \textbf{1.000} & --- \\
Matched-coverage raw (top 71) & 71 & 0.676 & 14 (Jaccard 0.110) \\
Top 10\% raw & 40 & 0.775 & --- \\
Top 10\% smoothed & 40 & 0.825 & --- \\
\bottomrule
\end{tabular}
\end{table}

A gamma sweep and a linear-scorer ablation confirm that the abstention rule is
not a smoothing artifact: the boundary cut remains the strongest zero-fit
selector, and the zero-cut subset is unchanged
(Appendix~\ref{app:gamma}, \ref{app:raw-smoothed}).

Changing the scorer or encoder changes the retrieval scores, and hence can change path resolution and the continuous edge weights; the abstention rule is then recomputed on the resulting selected-support field and graph. Table~\ref{tab:transfer} reports boundary cut performance under three configurations on the 400-query MAB multi-hop set: empirically, the same 71 queries have a zero boundary cut in all three, and all 71 are correct. This is a structural result about the selected supports and their boundaries, not exact invariance of continuous neural scores. Figure~\ref{fig:aupcc-summary} shows the per-context-length aggregate.

\begin{table}[tbp]
\centering
\caption{Transfer across scorers and encoders (multi-hop, 400 queries each). Same 71 zero-energy queries, same 100\% precision in all configurations.}
\label{tab:transfer}
\begin{tabular}{llcccc}
\toprule
Scorer & Encoder & $d$ & Coh.\ AUPCC & Native smoothed & Gap \\
\midrule
$(\mathbf{q}^\top\mathbf{v})^2$ & Qwen   & 128 & 0.785 & 0.677 & +0.108 \\
$\mathbf{q}^\top\mathbf{v}$     & Qwen   & 128 & 0.775 & 0.690 & +0.085 \\
$(\mathbf{q}^\top\mathbf{v})^2$ & MiniLM & 384 & 0.788 & 0.680 & +0.108 \\
\bottomrule
\end{tabular}
\end{table}

\paragraph{What produces the zero-cut cases}
\label{sec:zero-mechanism}
The zero-cut rule identifies an empirically high-precision structural-closure regime. Closure is determined entirely by the upstream resolver and neighborhood construction, without correctness labels. On MAB, this event selects 71 of 400 evaluations, all correct. Thus closure serves as a practical indicator of reliability in this pipeline, although it is not a logical certificate that the returned answer is true.

We audited the topology of the zero-cut neighborhoods. The original
run retained per-query scalar results but not graph snapshots, so the
audit uses a regeneration with the pinned extraction model
(gpt-4.1-mini-2025-04-14) and the original retrieval configuration,
which yields 70 zero-cut queries, 67 overlapping the reported 71.
None is trivial or disconnected: every regenerated zero-cut
neighborhood is a single small connected component with internal edges. The mechanism is hard selection: the typed one-hop neighborhood induced by the resolved path contains only the
selected path nodes in all 70 cases, but not in any of the 330 non-zero cases, so
the indicator field is constant and its cut boundary is empty. This closure is produced upstream by neighborhood construction and path resolution; the abstention rule detects it, it does not create it. The zeros mean no unselected candidate
remains in the constructed local neighborhood---not that no competing evidence exists in the world, and not equality of continuous
retrieval scores: a continuous score field over the same
neighborhoods yields no exact zeros in 400 queries, and the zero set
contains no duplicate candidates. Base accuracy is 56.8\% among the
396/400 regenerated queries with a non-trivial connected component (the
4 trivial queries are all incorrect), versus 70/70 in the zero-cut
subset, so zero cut marks a high-precision region within the
non-trivial population.

\paragraph{Controlling for unresolved retrieval}
Boundary cut ranks unresolved (\textsc{not\_found}) queries last through its cut
sentinel, whereas native score baselines retain their (overlapping) scores, so the
native full-population comparison partly reflects unresolved-status handling. We
report three comparisons grouped by exact question (10{,}000 replicates, unit =
exact question text): (A) \emph{native full} (Table~\ref{tab:main}), boundary cut $0.785$
vs.\ smoothed $0.677$, with $\Delta=+0.108$ (95\% CI $[+0.062, +0.156]$); (B) \emph{status-matched
full}, a post-hoc diagnostic that gives each score baseline the same unresolved-last
placement while leaving resolved-row scores untouched, boundary cut $0.785$ vs.\ smoothed
$0.736$, $\Delta=+0.049$ ($[+0.008, +0.092]$); and (C) \emph{resolved-only} ($N{=}320$),
boundary cut $0.823$ vs.\ smoothed $0.762$, $\Delta=+0.061$ ($[+0.010, +0.114]$). The advantage is not a sentinel
artifact: it remains positive with intervals excluding zero under both status matching and
resolved-only evaluation. On the regenerated population used for the structural controls and QPP comparison, the same boundary-cut-versus-smoothed difference is $+0.027$ with a CI including zero (Table~\ref{tab:stage3-qpp-audit}), so the aggregate score-baseline advantage is population-sensitive; the regenerated analysis retains the closure regime and an advantage over all four QPP adaptations, although QPP was evaluated only on that population. The stricter resolved non-closure analysis below nevertheless
removes the raw-boundary advantage.

Several analyses intentionally use different preserved or cached populations; their
absolute AUPCC values answer different questions and are not interchangeable with the
end-to-end $0.785$ headline.

\begin{table}[tbp]
\centering
\small
\caption{Provenance of MAB boundary-cut AUPCC values across analyses.}
\label{tab:mab-provenance}
\begin{tabular}{lll}
\toprule
AUPCC & Analysis & Relation to 0.785 \\
\midrule
0.785 & Original end-to-end MAB (pooled 400, d128) & headline \\
0.782 & Regenerated mechanism audit (400; 70 zero-energy) & different regeneration \\
0.765 & Graph ablation (cached candidates/path; graph varied) & within-table only \\
0.765--0.766 & Gamma sweep (separate cached harness; path re-resolved per $\gamma$) & within-sweep only \\
0.750 & Edge perturbation (own cached seed/protocol) & perturbation baseline \\
\bottomrule
\end{tabular}
\end{table}
The full-prefix evaluator gives $0.7849$ on the same headline run; Table~\ref{tab:main}'s
50-point evaluator rounds to $0.785$.

\begin{figure}[t]
\centering
\includegraphics[width=5in]{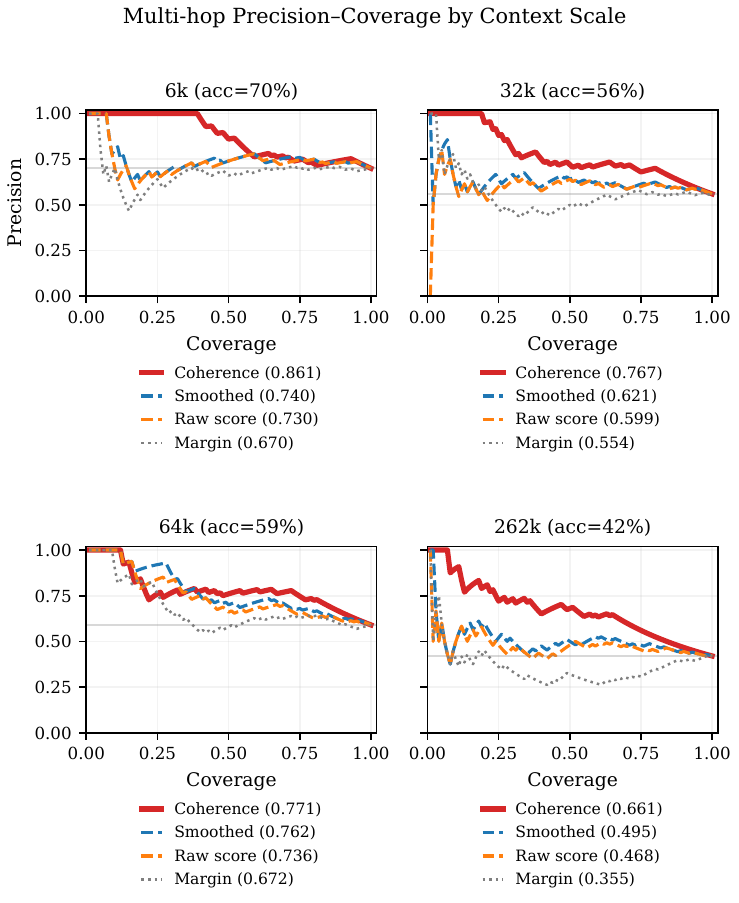}
\Description{Four multi-hop panels plot precision against coverage at 6k, 32k, 64k, and 262k context lengths. The boundary cut is compared with smoothed score, raw score, and margin. The boundary cut has the highest AUPCC in every panel, with respective values 0.861, 0.767, 0.771, and 0.661. Precision decreases overall as coverage increases.}
\caption{Multi-hop precision--coverage (6k/32k/64k/262k). The boundary cut (red) attains the highest AUPCC on all four multi-hop context lengths.}
\label{fig:mh-pc}
\end{figure}

\subsection{Structural Controls: Closure, Boundary Topology, and Resolution State}
\label{sec:structural-controls}

The zero-cut subset coincides exactly with structural closure $P=V$
(\S\ref{sec:zero-mechanism}). This raises a sharper question: outside that
binary closure, does the \emph{graded} boundary carry selective information,
and in particular do the continuous edge magnitudes matter beyond unweighted
boundary topology? Holding the constructed graph and the selected support fixed, we
compare the weighted boundary cut $E_w(P) = \sum_{i \in P,\, j \notin P} w_{ij}$
against structural controls: the unweighted boundary count
$E_1(P) = |\{(i,j) : i \in P,\, j \notin P,\, w_{ij} > 0\}|$, the selected
fraction $|P|/|V|$, the complement size $|V \setminus P|$, and the binary
closure indicator $\mathbf 1[P=V]$. Each is scored as a selective-prediction
signal with the same AUPCC evaluator. The verified regenerated population contains 374 distinct exact question texts across 400 evaluations; E1 and E1A resample exact-question groups. The analysis runs on the regenerated population of \S\ref{sec:zero-mechanism} (the only per-query artifact
that retains $V$, $P$, and edge weights), on which the weighted cut reproduces
AUPCC $0.7819$ (vs.\ the paper's $0.785$; the $-0.0031$ gap is the documented
70-vs-71 zero-count difference; Table~\ref{tab:mab-provenance}). Full definitions, hypotheses, and protocol are
in Appendix~\ref{app:e1}.

On the full 400-query set (Table~\ref{tab:mechanism-decomp}), the weighted cut (AUPCC $0.7819$) is statistically
indistinguishable from the unweighted boundary count ($0.7828$; paired
$\Delta = -0.0009$, 95\% CI $[-0.004, +0.003]$), the selected fraction
($0.7832$), and the complement size ($0.7811$). It does materially exceed the
binary closure indicator alone ($0.7199$; $\Delta = +0.0620$, 95\% CI
$[+0.030, +0.092]$): the graded boundary carries information beyond closure, but that information does not require the continuous weights.

Restricting to the $330$ queries with $P \neq V$ removes the exact closure
cases (base accuracy $0.4697$). The weighted cut retains selective value above
base (AUPCC $0.5837$, $+0.11$ over base) but again does not beat the unweighted
count ($0.5900$; paired $\Delta = -0.0064$, 95\% CI $[-0.020, +0.026]$).
This does not eliminate all apparent selective
signal, but this experiment alone does not establish why that residual
separation occurs: the $P \neq V$ subset still mixes resolved and unresolved
(chain-broken) states, and we do not attribute the residual to boundary
topology per se.

These controls establish that the MAB advantage does not depend on continuous Laplacian weights. In this pipeline, inexpensive measures of selected-set topology recover essentially the same selective behavior. The practical implication is that the abstention statistic can be simplified when the graph structure supports it. This is an empirical finding about the MAB graph construction; the external results show that edge weights matter in other regimes (\S\ref{sec:cross-benchmark}).

A normalized variant (the boundary-weight fraction) separates more strongly, but our preregistered controls did not isolate normalization from the weights themselves, so we develop it as a hypothesis below.

\paragraph{Resolution state: where the residual comes from}
\label{sec:resolution-state}
The unresolved set is the $84$ queries whose multi-hop chain broke (this
regenerated structural-control population has $84$ unresolved, versus $80$ in the
original headline run; see \S\ref{sec:setup}). Within this regenerated
analysis all $84$ are incorrect and receive the structural-control family's
worst-confidence sentinel, so every \emph{structural-control} signal ranks them
last and gains an easy failure-state distinction; this concerns the boundary
signals of this analysis, not the headline score baselines. Of the $84$, $80$
carry a partial resolved prefix and $4$ have an empty path. Stratifying by resolution state, on the same population
and evaluator, isolates the effect.

Restricting to resolved non-closure queries (resolved and $P \neq V$; $N=246$,
base accuracy $0.6301$), the raw boundary signals fall to or below base
accuracy: the weighted cut reaches AUPCC $0.6000$ and the unweighted count
$0.6104$, both under $0.6301$. The apparent raw-boundary advantage outside
$P=V$ does not survive this stricter control; much of the earlier residual
separation is consistent with resolution-state composition (failure cases ranked
last) rather than a graded raw-boundary confidence signal among resolved
answers.

Normalizing the boundary by the local graph's total edge weight or count changes
the picture. The weighted normalized boundary (the boundary-weight fraction,
AUPCC $0.6968$) and its unweighted analogue (the boundary-edge fraction,
$E_1(P)/(\text{total positive edge count})$, AUPCC $0.7005$) both sit about
$+0.07$ above base accuracy. Weighted normalization improves the point estimate,
but its paired interval includes zero; the unweighted normalization interval excludes zero (weighted normalized vs.\ raw weighted $\Delta=+0.0968$, CI
$[-0.0002, +0.196]$; unweighted normalized vs.\ raw unweighted $\Delta=+0.0901$,
CI $[+0.007, +0.201]$; intervals are wide at $N=246$).

Continuous weights again add nothing: the weighted and unweighted normalized
fractions are statistically indistinguishable ($\Delta=-0.0037$, 95\% CI
$[-0.024, +0.016]$, the point estimate trivially favoring the unweighted
fraction). This generates a narrower, MAB-specific hypothesis, not an
established mechanism: \emph{relative boundary exposure}, the fraction of the
local graph that crosses the selected/unselected boundary, rather than raw cut
magnitude, may be the useful structural quantity. Because the unweighted
normalized boundary was motivated by the same data it is evaluated on, external
validation is required before any general claim, which the next subsection
supplies.

\begin{table}[htbp]
\centering\footnotesize
\caption{Boundary-signal AUPCC on the regenerated MAB multi-hop population: full ($N=400$) and resolved non-closure ($N=246$). Paired intervals in the accompanying analysis resample exact question texts.}
\label{tab:mechanism-decomp}
\begin{tabular}{ll}
\toprule
Population / signal & AUPCC \\
\midrule
\multicolumn{2}{l}{\emph{Full ($N=400$)}} \\
Weighted cut & 0.782 \\
Unweighted cut & 0.783 \\
\midrule
\multicolumn{2}{l}{\emph{Resolved non-closure ($N=246$)}} \\
Weighted cut & 0.600 \\
Unweighted cut & 0.610 \\
Boundary-weight fraction & 0.697 \\
Boundary-edge fraction & 0.701 \\
\bottomrule
\end{tabular}

\end{table}

\subsection{Cross-Benchmark Test of the Normalization Hypothesis}
\label{sec:cross-benchmark}

We test the MAB-specific normalization hypothesis of \S\ref{sec:structural-controls} outside MAB, on existing structured-retrieval pipelines with
their own graphs, selectors, and correctness definitions (configurations in Table~\ref{tab:external-config}). This is an
external-validation test, not a further within-MAB optimization.

Confirmatory interpretation is restricted to benchmark lanes whose original
published result we could reproduce exactly. LongMemEval temporal and semantic
both reproduced (boundary cut AUPCC $0.7356$ and $0.6818$; gate PASS), as did
MetaQA-3hop (reproduced boundary cut$-$raw $\Delta=+0.0198$ vs.\ published
$+0.023$; PASS). CoDEx-S failed its reproduction gate (the existing ComplEx
training is unseeded and the published initialization is unrecoverable), so it
is reported as gate-failed exploratory only (Appendix~\ref{app:e2}). The MAB
single-hop and LoCoMo lanes were provenance-blocked (Appendix~\ref{app:e2}) and
are not interpreted.

\begin{table}[tbp]
\centering
\footnotesize
\caption{External-benchmark evaluation configurations and reproduction status.}
\label{tab:external-config}
\begin{tabular}{lccll}
\toprule
Benchmark & $k$ & $\gamma$ & Retrieval / graph & Reproduction \\
\midrule
LongMemEval temporal & 5 & 0.100 & session chain, continuous & reproduced \\
LongMemEval semantic & 5 & 0.100 & mutual-kNN(5), continuous & reproduced \\
MetaQA-3hop & 10 & 0.100 & typed answer-relation + type filter & reproduced \\
CoDEx-S & 10 & 0.100 & co-occurrence / typed mask & exploratory \\
LoCoMo & 5 & 0.100 & temporal chain & blocked \\
\bottomrule
\end{tabular}
\end{table}

The normalization improvement seen on MAB does not replicate. On the LongMemEval
temporal graph the normalized unweighted boundary is no better than the raw
count (AUPCC $0.7363$ vs.\ $0.7451$; $\Delta=-0.0088$, 95\% CI
$[-0.026, +0.015]$), and on MetaQA-3hop it is numerically harmful ($0.2831$ vs.\
$0.3238$; $\Delta=-0.0407$, CI $[-0.095, +0.022]$, worse on every predefined
answer-relation). Across the two confirmatory external families the normalized
boundary never beats the raw count (0/2 positive point estimates).

The LongMemEval semantic graph behaves differently. There the raw weighted cut
is weak ($0.6818$) while the raw unweighted count is strong ($0.7793$), and
among the normalized signals the weighted normalized boundary ($0.8528$)
materially exceeds the unweighted normalized boundary ($0.7744$):
$\Delta=+0.0785$, 95\% CI $[+0.048, +0.111]$. The MAB finding that continuous
weights are unnecessary does not hold universally: continuous
weights can
matter in at least one external graph regime. Task-relevant edge-weight semantics (here, relevance-weighted similarity) is a plausible, regime-specific explanation; this interpretation is exploratory and post-hoc.

No single raw-versus-normalized or weighted-versus-unweighted rule generalized
across the evaluated external regimes: normalization helped only on MAB, was
null on the temporal graph, harmful on MetaQA, and on the semantic graph the
advantage came from weights rather than from normalization. Performance depends
on the interaction among candidate-space structure, graph construction,
normalization, and edge-weight semantics. The result is a regime-dependent structural diagnostic, not a universal replacement for the raw boundary cut.

\subsection{Typed Structure, Not Generic Similarity, Drives the Signal}
\label{sec:graph-ablation}

This ablation asks which \emph{edge set} (graph construction) works, holding
candidates and scores fixed; it is distinct from the structural controls of
\S\ref{sec:structural-controls}, which hold the graph fixed and ask whether
continuous edge \emph{magnitudes} matter beyond unweighted boundary topology.

Within this cached ablation (which holds the resolved path $P$ and scores fixed
and rebuilds only the edge weights) the typed graph gives AUPCC $0.765$, compared
with $0.716$ for semantic kNN and $0.748$ for the best hybrid construction; full graph-ablation results are in
Table~\ref{tab:graph-ablation-app}. The typed
graph is also the only variant that preserves the perfect zero-cut subset:
semantic kNN produces more zero-cut queries, but seven are wrong, reducing its
zero-cut precision to $91.7\%$. The result supports
Remark~\ref{rem:boundary}: generic semantic edges add
disagreement without adding correctness information, so the typed mask is
valuable because of what it excludes. The boundary cut does not need a
particular encoder or scorer, but it does need graph edges aligned with
answer support (robustness to edge corruption in Appendix~\ref{app:robustness}).

Mean boundary cut separates correct from incorrect answers most sharply for the typed graph (mean $E$ = 13.25 on correct vs.\ 29.48 on incorrect), and least for semantic kNN (2.28 vs.\ 4.02), with the hybrids intermediate ($\alpha{=}0.25$: 5.03 vs.\ 9.91; $\alpha{=}0.50$: 7.77 vs.\ 15.79; $\alpha{=}0.75$: 10.51 vs.\ 21.68).

\begin{table}[tbp]
\centering
\caption{Full MAB graph ablation on cached local neighborhoods (multi-hop).}
\label{tab:graph-ablation-app}
\begin{tabular}{lcccccc}
\toprule
Variant & Agg. & 6k & 32k & 64k & 262k & Zero-energy precision \\
\midrule
Typed entity--relation & \textbf{0.765} & \textbf{0.866} & \textbf{0.743} & \textbf{0.754} & 0.592 & \textbf{1.000} \\
Semantic kNN & 0.716 & 0.827 & 0.683 & 0.655 & \textbf{0.598} & 0.917 \\
Hybrid ($\alpha=0.250$) & 0.745 & 0.855 & 0.722 & 0.723 & 0.567 & 0.986 \\
Hybrid ($\alpha=0.500$) & 0.747 & 0.858 & 0.722 & 0.738 & 0.560 & 0.986 \\
Hybrid ($\alpha=0.750$) & 0.748 & 0.857 & 0.724 & 0.737 & 0.564 & 0.986 \\
\bottomrule
\end{tabular}
\end{table}

\begin{figure}[tbp]
\centering
\includegraphics[width=5in]{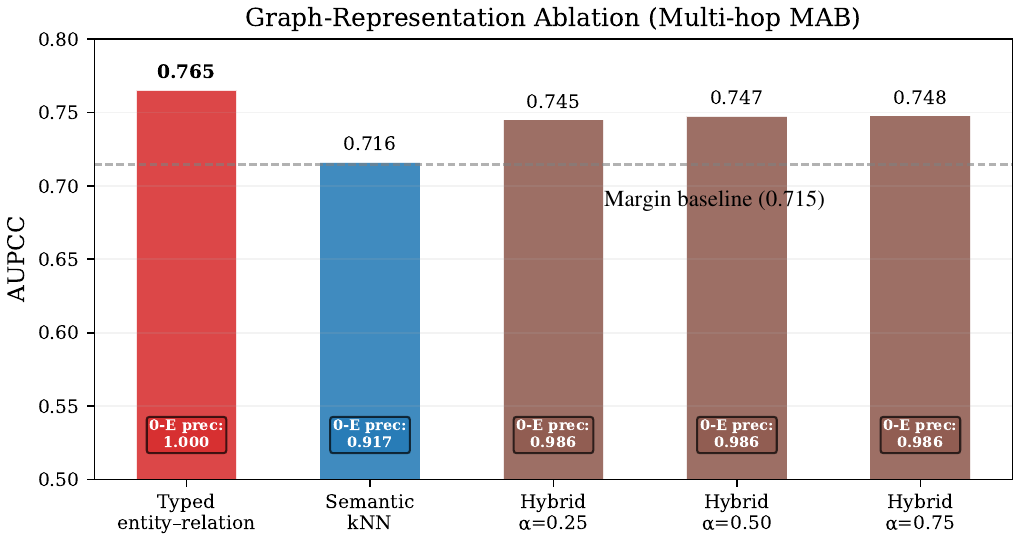}
\Description{Five bars show multi-hop AUPCC on cached neighborhoods: typed entity-relation 0.765, semantic kNN 0.716, and hybrid graphs 0.745, 0.747, and 0.748. Labels inside the bars show zero-cut precision of 1.000, 0.917, and 0.986 for each hybrid. A horizontal dashed line marks the margin baseline.}
\caption{Graph-representation ablation on cached multi-hop MAB neighborhoods. Results are computed post hoc on cached local neighborhoods and are not directly comparable to the end-to-end AUPCC values in Tables~\ref{tab:main} and~\ref{tab:learned}. Typed entity--relation structure outperforms semantic kNN and hybrid alternatives. The typed graph is the only variant with a perfect zero-cut subset (71 queries, 100\% precision). Adding semantic edges dilutes the signal, consistent with Remark~\ref{rem:boundary}.}
\label{fig:graph-ablation-bar}
\end{figure}

\subsection{Support Compatibility Is Relation-Level, Not Dataset-Level}
\label{sec:per-relation-replication}
On LongMemEval, LoCoMo, and CoDEx-S, the boundary cut underperforms score
baselines at the dataset level (Table~\ref{tab:negative-transfer}).
Per-relation analysis shows a finer pattern. On CoDEx-S and
MetaQA-3hop, each dataset contains both support-compatible and
support-incompatible relations, so the aggregate result reflects a
mixture rather than a single dataset-level verdict. Neither dataset provides a positive aggregate benchmark for the method; the observed variation is relation-dependent. The
$R_{\mathrm{median}}$ gate correctly rejects both at the dataset level
(Appendix~\ref{app:gate}). Relation-level behavior is heterogeneous and we
treat it as exploratory: on the reproduced MetaQA evaluation only
\texttt{in\_language} has a paired bootstrap interval excluding zero, and the
CoDEx-S per-relation results are artifact-conditional (their exact historical
per-example predictions were not persisted). Illustrative relation-level
results in both directions, with paired bootstrap intervals, are in
Tables~\ref{tab:codex_per_relation} and~\ref{tab:metaqa_per_relation}. This secondary diagnostic consumes correctness labels; its threshold was chosen on the original evaluated families, not independently validated as a general predictive criterion.

\begin{table}[tbp]
\centering
\caption{Boundary-condition experiments. The boundary cut fails when the graph does not encode support-compatible structure.}
\label{tab:negative-transfer}
\begin{tabular}{lccc}
\toprule
Benchmark / graph & Coherence & Best comparator & Diagnosis \\
\midrule
LongMemEval / temporal & 0.736 & 0.887 & temporal adjacency \\
LongMemEval / semantic kNN & 0.682 & 0.890 & generic similarity \\
LoCoMo / temporal & 0.286 & 0.430 & temporal adjacency \\
CoDEx-S / typed entity--relation & 0.373 & 0.543 & heterogeneous, App.~\ref{app:per_relation} \\
\bottomrule
\end{tabular}
\end{table}

The negative-transfer results clarify the conditions under which graph structure provides useful confidence information. In these experiments, misaligned graph edges yield weak or misleading signals, while the positive MAB result is associated with edges that connect candidates supporting the same answer.

As a practical diagnostic, we compute three cheap metrics across six graph families to characterize when the boundary cut is useful (Table~\ref{tab:probe}): $R_\text{median}$, the ratio of median cut on wrong vs.\ correct answers (higher means better separation); AUROC of $-E$ as a correctness discriminator; and AUPCC. This probe is not a prospective predictor of when the boundary cut will work (it requires ground-truth labels), but it provides a useful post-hoc sanity check for new graph constructions. This secondary diagnostic consumes correctness labels; its threshold was chosen on the original evaluated families, not independently validated as a general predictive criterion.

\begin{table}[tbp]
\centering
\caption{Support-compatibility probe across six graph families. High $R_\text{median}$ and AUROC describe observed correctness separation; values near 1.0 or below 0.5 indicate negative transfer. This secondary diagnostic consumes correctness labels; its threshold was chosen on the original evaluated families, not independently validated as a general predictive criterion. ZE denotes zero-energy.}
\label{tab:probe}
\begin{tabular}{lcccccc}
\toprule
Graph family & $N$ & Acc & $R_\text{median}$ & AUROC & AUPCC & ZE prec \\
\midrule
MAB typed         & 400  & 0.560 & 8.150  & 0.755 & 0.765 & 1.000 \\
MAB semantic kNN  & 400  & 0.560 & 2.750  & 0.692 & 0.716 & 0.917 \\
LongMemEval temp. & 500  & 0.740 & 1.010  & 0.488 & 0.736 & --- \\
LongMemEval kNN   & 500  & 0.750 & 0.510  & 0.414 & 0.682 & --- \\
LoCoMo temporal   & 1986 & 0.270 & 1.010  & 0.517 & 0.286 & --- \\
CoDEx-S co-occ.   & 1828 & 0.330 & 0.670  & 0.332 & 0.215 & 0.083 \\
\bottomrule
\end{tabular}
\end{table}

The pattern is clear: MAB's typed entity--relation graph produces large cut separation ($R_\text{median}{=}8.15$, AUROC $0.755$) and a perfect zero-cut subset. Temporal adjacency produces $R_\text{median}{\approx}1$ (no separation). Semantic kNN and co-occurrence show intermediate or inverted ratios. These results are consistent with the support-compatibility hypothesis: the graph must connect candidates that should agree under the current query. This secondary diagnostic consumes correctness labels; its threshold was chosen on the original evaluated families, not independently validated as a general predictive criterion.

\begin{table}[tbp]
\centering
\caption{Per-relation boundary-cut performance on CoDEx-S with typed entity--relation mask. $R_{\mathrm{median}}$: ratio of median cut on incorrect vs.\ correct answers (reported as N/A when both medians are zero). Relations ordered by boundary-cut AUPCC. This secondary diagnostic consumes correctness labels; its threshold was chosen on the original evaluated families, not independently validated as a general predictive criterion.}
\label{tab:codex_per_relation}
\small
\begin{tabular}{llrrrrrrr}
\toprule
PID & Label & $N$ & Acc & AUPCC$_{\mathrm{coh}}$ & AUPCC$_{\mathrm{raw}}$ & $R_{\mathrm{med}}$ & ZE prec & ZE $n$ \\
\midrule
P101  & field of work       &  20 & 0.800 & 0.766 & 0.896 &  1.400 &   ---  &   0 \\
P463  & member of           & 270 & 0.580 & 0.700 & 0.624 & 52.000 & 0.764  &  89 \\
P1303 & instrument          &  71 & 0.560 & 0.577 & 0.719 &  1.470 & 0.600  &  20 \\
P27   & citizenship         & 105 & 0.600 & 0.485 & 0.795 &  0.690 & 0.000  &   2 \\
P530  & diplomatic relation & 287 & 0.240 & 0.256 & 0.390 &  N/A  & 0.244  & 287 \\
P106  & occupation          & 593 & 0.250 & 0.215 & 0.313 &  0.880 & 1.000  &   1 \\
P136  & genre               & 102 & 0.220 & 0.178 & 0.268 &  0.750 & 0.000  &   4 \\
\bottomrule
\end{tabular}
\end{table}

{\footnotesize \emph{Note:} For P530, the typed mask produces disconnected local
subgraphs, so zero cut is structural rather than informative.\par}

\begin{table}[tbp]
\centering
\caption{Illustrative relation-level results on MetaQA-3hop ($n{=}250$ stratified),
boundary cut vs.\ raw score, recomputed from the persisted per-query evaluation
artifact. Both directions shown; selection is illustrative, not statistically
representative. Intervals are paired bootstrap 95\% CIs using the per-query
sampling unit (10{,}000 replicates, seed 42), descriptive and uncorrected for
multiple comparisons; small-$N$ relations have wide intervals. Only
\texttt{in\_language} has an interval excluding zero.}
\label{tab:metaqa_per_relation}
\small
\begin{tabular}{lrrrrl}
\toprule
Answer relation & $N$ & Base acc & AUPCC$_{\mathrm{coh}}$ & AUPCC$_{\mathrm{raw}}$ & $\Delta$ (95\% CI) \\
\midrule
\texttt{in\_language}    & 46 & 0.435 & 0.546 & 0.317 & $+0.230$ $[+0.030, +0.407]$ \\
\texttt{release\_year}   & 41 & 0.341 & 0.367 & 0.306 & $+0.062$ $[-0.120, +0.246]$ \\
\texttt{has\_genre}      & 56 & 0.304 & 0.352 & 0.326 & $+0.026$ $[-0.128, +0.157]$ \\
\texttt{starred\_actors} & 31 & 0.226 & 0.168 & 0.141 & $+0.027$ $[-0.137, +0.192]$ \\
\texttt{directed\_by}    & 39 & 0.282 & 0.287 & 0.333 & $-0.046$ $[-0.292, +0.160]$ \\
\texttt{written\_by}     & 37 & 0.297 & 0.209 & 0.271 & $-0.062$ $[-0.256, +0.137]$ \\
\bottomrule
\end{tabular}
\end{table}

The pattern tracks candidate-space structure: closed-vocabulary or structurally dense answer types (organizational membership, languages, release years) yield small, connected typed neighborhoods in which incorrect answers cross a non-empty boundary, whereas open-vocabulary or sparse types (person names; arbitrary-tail relations such as P530) yield high-cardinality, sparse or disconnected neighborhoods where the cut carries little information. On MetaQA the answer-type candidate filter performs most of this narrowing and the typed edge mask is nearly a no-op. Tables~\ref{tab:codex_per_relation} and~\ref{tab:metaqa_per_relation} give the per-relation results; Appendix~\ref{app:per_relation} details the mechanism and candidate-filter versus edge-mask comparison.

\subsection{Comparison with Learned Confidence}
\begin{table}[htbp]
\centering\footnotesize
\caption{Learned and training-free confidence on headline MAB multi-hop ($N=400$), using the reconstructed mapping (Appendix~\ref{app:cross-config}). GBT and LR use five-fold cross-validation grouped by exact question; paired uncertainty resamples exact question texts.}
\label{tab:learned}
\begin{tabular}{lll}
\toprule
Signal & AUPCC & AUROC \\
\midrule
GBT & 0.814 & 0.889 \\
Coherence & 0.785 & 0.807 \\
Smoothed score & 0.677 & 0.645 \\
Raw score & 0.653 & 0.611 \\
LR & 0.648 & 0.620 \\
Score margin & 0.547 & 0.413 \\
\bottomrule
\end{tabular}
\par\smallskip\begin{minipage}{0.95\linewidth}\scriptsize Native row order; per-fold StandardScaler. CIs condition on fixed out-of-fold predictions.\end{minipage}
\end{table}

A shallow gradient-boosted tree ensemble (GBT; depth~3, 50~trees) trained via five-fold CV grouped by exact question on five retrieval-time features achieves AUPCC $0.814$, versus the boundary cut's $0.785$ (Table~\ref{tab:learned}). Under the reconstructed mapping (Appendix~\ref{app:cross-config}), the main native-order, training-fold-standardized protocol gives boundary cut minus GBT $\Delta=-0.029$ (95\% CI $[-0.062, +0.005]$), so the difference is not statistically resolved. L2-regularized logistic regression achieves $0.648$. A separate native-order unscaled sensitivity yields GBT AUPCC $0.812$ and boundary cut minus GBT $\Delta=-0.027$ (95\% CI $[-0.060, +0.008]$). This sensitivity is distinct from the main estimate; both protocols and their grouping history are documented in Appendix~\ref{app:cross-config}.

\begin{figure}[tbp]
\centering
\includegraphics[width=5in]{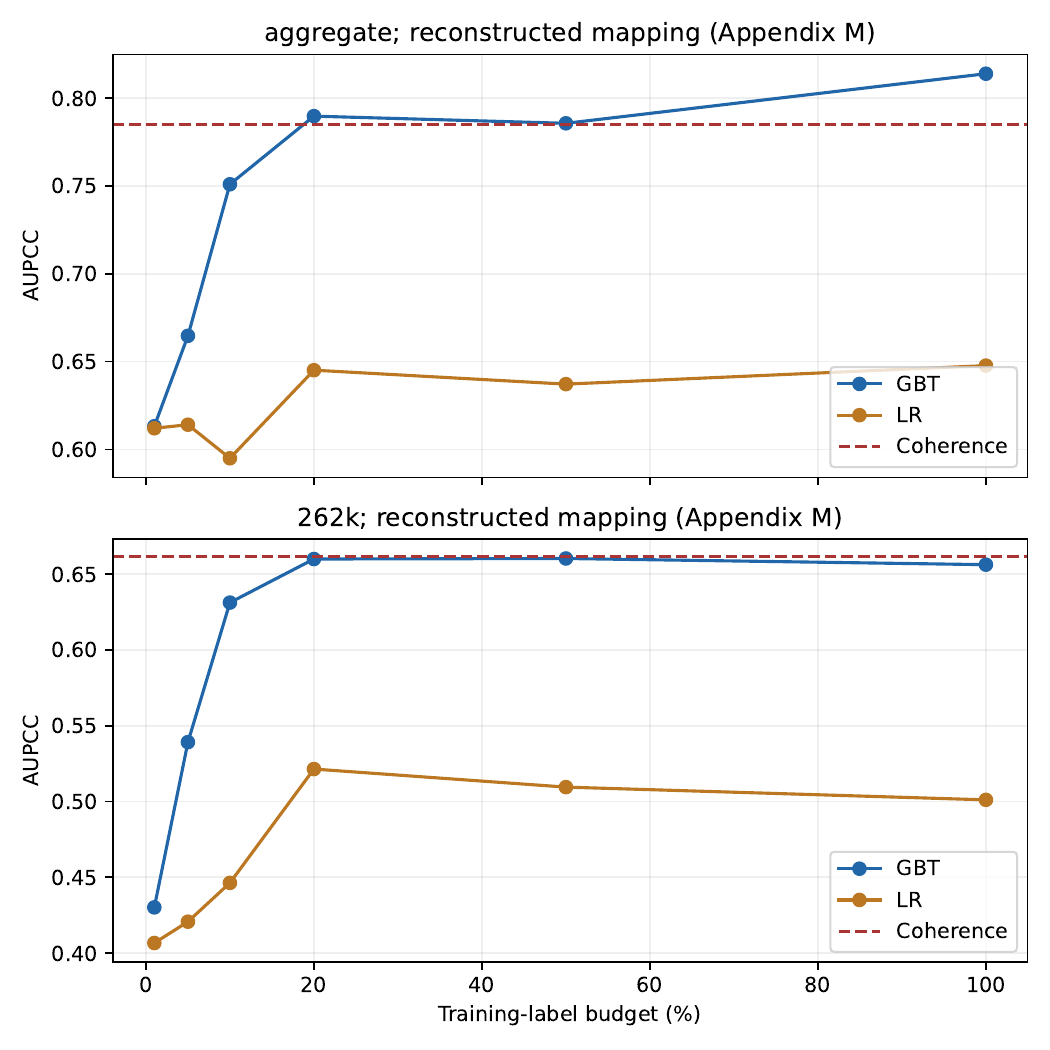}
\Description{Two panels plot AUPCC against training-label fraction for gradient-boosted and logistic-regression selectors, with a horizontal boundary cut baseline. Aggregate GBT first exceeds the boundary cut at the 20 percent label budget. On the 262k split, GBT remains below the boundary cut at every plotted budget. Logistic regression is lower than the boundary cut in both panels.}
\caption{Label-efficiency curve on multi-hop MAB. Boundary cut (zero-fit, horizontal line) vs.\ GBT trained on subsampled labels. Top: aggregate AUPCC. Bottom: 262k point estimates. GBT first exceeds the boundary cut in aggregate by point estimate at 20\% labels; paired intervals include zero at 20\%, 50\%, and 100\%. Question identity uses the reconstructed mapping (Appendix~\ref{app:cross-config}).}
\label{fig:label-efficiency}
\end{figure}

\paragraph{Label-efficiency table}
Table~\ref{tab:labeleff} reports GBT performance trained on subsamples of training data (1\%--100\%) within each CV fold grouped by exact question, alongside the constant zero-fit boundary cut baseline. GBT first exceeds the boundary cut by point estimate at 20\% labels; the paired aggregate intervals include zero at 20\%, 50\%, and 100\%. Figure~\ref{fig:label-efficiency} plots the aggregate curve and the 262k point estimates. All headline-derived results use the reconstructed mapping (Appendix~\ref{app:cross-config}).

\begin{table}[htbp]
\centering\footnotesize
\caption{Label-budget comparison on headline MAB multi-hop ($N=400$ overall; $N=100$ at 262k). Budgets are percentages of training rows within folds grouped by exact question; paired intervals resample exact question texts. Question identity uses the reconstructed mapping (Appendix~\ref{app:cross-config}).}
\label{tab:labeleff}
\begin{tabular}{lllll}
\toprule
Labels (\%) & GBT (all) & Coherence (all) & GBT (262k) & Coherence (262k) \\
\midrule
1 & 0.613 & 0.785 & 0.430 & 0.661 \\
5 & 0.665 & 0.785 & 0.539 & 0.661 \\
10 & 0.751 & 0.785 & 0.631 & 0.661 \\
20 & 0.790 & 0.785 & 0.660 & 0.661 \\
50 & 0.786 & 0.785 & 0.660 & 0.661 \\
100 & 0.814 & 0.785 & 0.656 & 0.661 \\
\bottomrule
\end{tabular}
\end{table}

AUROC and cross-scale generalization details are in Appendix~\ref{app:learned-confidence-extras}.

\paragraph{Setup and maintenance tradeoff}
Supervised selection attains a higher in-distribution point estimate when correctness labels and suitable confidence features are available, though the paired difference is not statistically resolved on 400 queries. Graph-boundary abstention offers a different operating point: a training-free rule computed from deployment-time evidence structure, without maintaining a separate learned confidence model.
\label{sec:cross-config-robustness}
We test how the GBT point-estimate comparison changes with the retrieval stack. We hold the 400 multi-hop MAB queries fixed and use three configurations: \textsc{base} (Qwen, $(q^\top v)^2$), \textsc{minilm} (MiniLM, $(q^\top v)^2$; encoder shift), and \textsc{linear} (Qwen, $q^\top v$; scorer shift). For each (source, target) pair we train GBT on source-config features+labels under 5-fold CV grouped by exact question text, and evaluate on target-config features against target-config labels (no query overlap, no target labels used for fitting). Tables~\ref{tab:cross-config-transfer} and~\ref{tab:cross-config-bootstrap} give the transfer matrix and per-cell bootstrap confidence intervals; Appendix~\ref{app:cross-config} gives the AUROC matrix and feature-scale distributions.

Under cross-validation grouped by exact question with the reconstructed mapping (Appendix~\ref{app:cross-config}) (Table~\ref{tab:cross-config-bootstrap}), the MINILM and LINEAR matched-stack comparisons favor GBT, while the BASE matched-stack difference is unresolved. Among off-diagonal cells, BASE$\to$LINEAR favors GBT; MINILM$\to$BASE and LINEAR$\to$MINILM favor the boundary cut; the remaining three differences are unresolved. In particular, BASE$\to$BASE has a 95\% CI of $[-0.0000, +0.065]$, MINILM$\to$BASE $[-0.097, -0.012]$, MINILM$\to$MINILM $[+0.005, +0.073]$, and LINEAR$\to$BASE $[-0.084, +0.004]$. Feature-scale shift remains a plausible explanation for transfer degradation.

\begin{table}[htbp]
\centering\footnotesize
\caption{GBT transfer AUPCC across retrieval configurations on headline MAB multi-hop ($N=400$ per target). Rows identify training configurations and columns deployment configurations. Folds and paired resampling use exact question texts under the reconstructed mapping (Appendix~\ref{app:cross-config}).}
\label{tab:cross-config-transfer}
\begin{tabular}{llll}
\toprule
Train / test & \textsc{base} & \textsc{minilm} & \textsc{linear} \\
\midrule
\textsc{base} & 0.818 & 0.771 & 0.825 \\
\textsc{minilm} & 0.739 & 0.828 & 0.763 \\
\textsc{linear} & 0.742 & 0.696 & 0.838 \\
\bottomrule
\end{tabular}
\end{table}

\paragraph{Bootstrap 95\% confidence intervals (10{,}000 resamples grouped by exact question)}
Table~\ref{tab:cross-config-bootstrap} compares each GBT cell against the target-configuration boundary cut using 10{,}000 paired exact-question bootstrap replicates, seed 42, conditional on the fixed out-of-fold predictions. The archived correction used 2{,}000 replicates despite the earlier manuscript's 10{,}000 description; the repaired intervals use 10{,}000 throughout.

\begin{table}[htbp]
\centering\footnotesize
\caption{Paired AUPCC differences between transferred GBT and the target-configuration boundary cut on headline MAB multi-hop ($N=400$ per target). Intervals use 10,000 exact-question bootstrap replicates, conditional on out-of-fold predictions, under the reconstructed mapping (Appendix~\ref{app:cross-config}).}
\label{tab:cross-config-bootstrap}
\begin{tabular}{llllll}
\toprule
Cell & GBT & Coherence & $\Delta$ & 95\% CI & Direction \\
\midrule
\textsc{base}$\to$\textsc{base} & 0.818 & 0.785 & 0.033 & [-0.0000, +0.065] & unresolved \\
\textsc{base}$\to$\textsc{minilm} & 0.771 & 0.788 & -0.017 & [-0.050, +0.024] & unresolved \\
\textsc{base}$\to$\textsc{linear} & 0.825 & 0.775 & 0.050 & [+0.016, +0.087] & positive \\
\textsc{minilm}$\to$\textsc{base} & 0.739 & 0.785 & -0.046 & [-0.097, -0.012] & negative \\
\textsc{minilm}$\to$\textsc{minilm} & 0.828 & 0.788 & 0.041 & [+0.005, +0.073] & positive \\
\textsc{minilm}$\to$\textsc{linear} & 0.763 & 0.775 & -0.012 & [-0.059, +0.029] & unresolved \\
\textsc{linear}$\to$\textsc{base} & 0.742 & 0.785 & -0.043 & [-0.084, +0.004] & unresolved \\
\textsc{linear}$\to$\textsc{minilm} & 0.696 & 0.788 & -0.092 & [-0.132, -0.034] & negative \\
\textsc{linear}$\to$\textsc{linear} & 0.838 & 0.775 & 0.063 & [+0.031, +0.096] & positive \\
\bottomrule
\end{tabular}
\par\smallskip\begin{minipage}{0.95\linewidth}\scriptsize Values are rounded to three decimals; a fourth decimal preserves the sign of near-zero CI endpoints. The displayed $-0.0000$ is negative at full precision, so that interval includes zero. Significance is determined before rounding.\end{minipage}
\end{table}

A reconstructed post-hoc population-rank diagnostic substantially mitigates transfer degradation: LINEAR$\to$MINILM increases from $0.696$ to $0.771$. This result concerns ranks across queries within each configuration, not within-query candidate ranks. The boundary cut is recomputed on the deployment path and remains perfectly preserved under encoder shift (BASE$\leftrightarrow$MINILM Pearson $\rho=1.000$) and partially preserved under scorer shift ($\rho=0.41$).

\paragraph{Reconstructed post-hoc percentile-rank diagnostic}
Table~\ref{tab:cross-config-pr-population-repaired} reports the archived population-rank transformation under repaired exact-question folds and the reconstructed mapping (Appendix~\ref{app:cross-config}). Each feature column is ranked across all 400 rows within its configuration before cross-validation, including held-out rows, with positional tie breaks. This is a transductive post-hoc diagnostic. Eight of nine cells reach or exceed the target-configuration boundary cut; the exception is LINEAR$\to$MINILM ($0.771$ versus $0.788$). Literal within-query candidate percentiles are unavailable because the headline features omit candidate-score vectors. The eight-of-nine population-rank result therefore does not establish a within-query normalization claim.

\begin{table}[htbp]
\centering\footnotesize
\caption{GBT transfer AUPCC after transductive population-rank normalization on headline MAB multi-hop ($N=400$ per configuration). Exact-question folds and paired resampling use the reconstructed mapping (Appendix~\ref{app:cross-config}). Feature ranks include held-out rows; this diagnostic is not within-query normalization.}
\label{tab:cross-config-pr-population-repaired}
\begin{tabular}{llll}
\toprule
Train / test & \textsc{base} & \textsc{minilm} & \textsc{linear} \\
\midrule
\textsc{base} & 0.827 & 0.805 & 0.824 \\
\textsc{minilm} & 0.808 & 0.820 & 0.792 \\
\textsc{linear} & 0.808 & 0.771 & 0.831 \\
\bottomrule
\end{tabular}
\par\smallskip\begin{minipage}{0.95\linewidth}\scriptsize Transductive per-feature ranks across 400 rows per configuration. This does not implement within-query candidate ranks.\end{minipage}
\end{table}

\paragraph{Search baselines}
We evaluated search methods only on multi-hop 6k (100 queries; Table~\ref{tab:search}); results at other scales may differ.

\begin{table}[tbp]
\centering
\caption{Search baselines on multi-hop 6k. Search does not improve accuracy in this setting; chain methods hurt.}
\label{tab:search}
\begin{tabular}{lcc}
\toprule
Method & Accuracy & $\Delta$ \\
\midrule
Baseline          & 0.660 & --- \\
Leaf MCTS         & 0.660 & +0.000 \\
Chain MCTS        & 0.580 & $-$0.080 \\
Path reranking    & 0.580 & $-$0.080 \\
\bottomrule
\end{tabular}
\end{table}

In this setting, the dominant failure mode is \emph{error compounding}: an incorrect first hop produces a corrupted query that drives later hops into an incorrect subtree with high confidence. The first hop is not the bottleneck (98/100 queries have a unique first-hop candidate, so first-hop reranking can affect at most $2\%$ of queries); the path-reranking and chain-MCTS lanes lose 8~points because their chain-aware greedy descent at \emph{subsequent} hops follows a cumulative-log-score signal that prefers locally-plausible but globally-incorrect paths over the deterministic smoothed-lookup choice. When the failure mode is local scoring ambiguity rather than incomplete exploration, abstention is the more effective use of post-retrieval compute.

\FloatBarrier
\paragraph{Score-distribution QPP comparison}
Table~\ref{tab:stage3-qpp-audit} reports the Stage 3 comparison on the regenerated population, using minimum attempted-hop aggregation. NQC-local, WIG-local, $\sigma_{\max}$, and $\sigma_{50\%}$ are local score-distribution adaptations, not corpus-score QPP evaluations. The boundary cut exceeds all four in AUPCC, with 95\% CIs grouped by exact question excluding zero in the status-matched, native, and resolved-only settings. All four adaptations underperform the paper's smoothed baseline by point estimate on these same rows. The best-QPP-minus-smoothed interval excludes zero in the primary status-matched and native settings, but includes zero in resolved-only evaluation; best-QPP selection is post hoc. The boundary cut's difference from smoothed is unresolved on this regenerated population. Formulas, provenance, and final-hop sensitivity results are available from the author.

\begin{table}[htbp]
\centering
\footnotesize
\setlength{\tabcolsep}{3pt}
\caption{Frozen regenerated population: minimum attempted-hop aggregation. NQC and WIG use a local-background score proxy. All scores are recomputed on the same rows.}
\label{tab:stage3-qpp-audit}
\begin{tabular}{lrrrrrr}
\toprule
 & \multicolumn{2}{c}{Status-matched (primary)} & \multicolumn{2}{c}{Native} & \multicolumn{2}{c}{Resolved-only} \\
Signal & AUPCC & AUROC & AUPCC & AUROC & AUPCC & AUROC \\
\midrule
Coherence & 0.782 & 0.810 & 0.782 & 0.810 & 0.822 & 0.635 \\
Raw score & 0.755 & 0.814 & 0.755 & 0.813 & 0.788 & 0.643 \\
Smoothed score & 0.754 & 0.815 & 0.754 & 0.814 & 0.787 & 0.643 \\
Smoothed full-vector margin & 0.666 & 0.729 & 0.593 & 0.559 & 0.675 & 0.479 \\
NQC-local & 0.669 & 0.730 & 0.558 & 0.520 & 0.679 & 0.481 \\
WIG-local & 0.699 & 0.742 & 0.543 & 0.495 & 0.719 & 0.504 \\
$\sigma_{\max}$ & 0.661 & 0.719 & 0.547 & 0.493 & 0.669 & 0.459 \\
$\sigma_{50\%}$ & 0.676 & 0.726 & 0.565 & 0.508 & 0.688 & 0.473 \\
\midrule
\multicolumn{7}{l}{Paired AUPCC differences [95\% bootstrap CI grouped by exact question]} \\
Coherence $-$ Raw score & \multicolumn{2}{c}{0.027 [-0.021, +0.076]} & \multicolumn{2}{c}{0.027 [-0.021, +0.076]} & \multicolumn{2}{c}{0.034 [-0.026, +0.094]} \\
Coherence $-$ Smoothed score & \multicolumn{2}{c}{0.027 [-0.019, +0.074]} & \multicolumn{2}{c}{0.028 [-0.019, +0.074]} & \multicolumn{2}{c}{0.035 [-0.024, +0.093]} \\
Coherence $-$ Smoothed full-vector margin & \multicolumn{2}{c}{0.116 [+0.073, +0.158]} & \multicolumn{2}{c}{0.189 [+0.143, +0.234]} & \multicolumn{2}{c}{0.147 [+0.092, +0.198]} \\
Coherence $-$ NQC-local & \multicolumn{2}{c}{0.113 [+0.065, +0.158]} & \multicolumn{2}{c}{0.224 [+0.169, +0.276]} & \multicolumn{2}{c}{0.143 [+0.083, +0.200]} \\
Coherence $-$ WIG-local & \multicolumn{2}{c}{0.083 [+0.037, +0.133]} & \multicolumn{2}{c}{0.238 [+0.170, +0.288]} & \multicolumn{2}{c}{0.103 [+0.047, +0.168]} \\
Coherence $-$ $\sigma_{\max}$ & \multicolumn{2}{c}{0.121 [+0.068, +0.170]} & \multicolumn{2}{c}{0.235 [+0.176, +0.290]} & \multicolumn{2}{c}{0.153 [+0.088, +0.216]} \\
Coherence $-$ $\sigma_{50\%}$ & \multicolumn{2}{c}{0.106 [+0.056, +0.153]} & \multicolumn{2}{c}{0.216 [+0.158, +0.270]} & \multicolumn{2}{c}{0.134 [+0.073, +0.193]} \\
\midrule
Best primary QPP & \multicolumn{2}{c}{WIG-local} & \multicolumn{2}{c}{$\sigma_{50\%}$} & \multicolumn{2}{c}{WIG-local} \\
Best QPP $-$ smoothed & \multicolumn{2}{c}{-0.055 [-0.114, -0.001]} & \multicolumn{2}{c}{-0.189 [-0.249, -0.128]} & \multicolumn{2}{c}{-0.069 [-0.146, +0.003]} \\
\bottomrule
\end{tabular}
\par\medskip
\begin{minipage}{0.980\linewidth}\footnotesize
$N=400$ for native/status-matched and $N=316$ for resolved-only. CIs use 10,000 paired bootstrap draws, seed 42, unit exact question text (verified from the frozen intermediate bundle). Best-QPP selection is post hoc; its intervals are descriptive. Smoothed scores use $(I+0.1L)^{-1}b$ at the frozen selected candidate. The margin uses the two highest smoothed scores over the full local vector. No historical headline baseline values are imported.
\end{minipage}
\end{table}

\FloatBarrier

\section{Related Work}
\label{sec:related}

Predicting retrieval quality without relevance judgments is the subject of query performance prediction (QPP). Post-retrieval predictors have relied on the distribution of retrieval scores or on the coherence of the top-ranked documents, and score-based predictors remain competitive with supervised ones~\citep{cronen2002qpp,zhou2006robustness,shtok2010decision,shtok2012drift,shtok2016reference,zamani2018neuralqpp,datta2022deepqpp,vlachou2024coherence}. Diaz~\citep{diaz2007spatial} predicts performance from the autocorrelation of scores across similar documents, the nearest precedent for graph-smoothness signals over retrieved items. QPP has also been extended to conversational search and to retrieval-augmented generation~\citep{faggioli2023geometric,salemi2024erag,huly2025rag}, and to dense retrieval under corpus growth~\citep{ghoshdastidar2026growth}. Abstention has been studied for reranking in black-box settings~\citep{gisserot2024abstention}, and calibration for knowledge-graph RAG has been proposed because such systems are often overconfident when retrieved subgraphs are incomplete~\citep{ren2026whentotrust}. Several QPP methods are also unsupervised. The distinction studied here is the boundary between evidence selected to support a returned answer and unselected candidates in a typed evidence graph, used as an answer-abstention signal. The name is unrelated to coherence-based QPP, which measures similarity within the top-ranked set.

Agent memory work~\citep{chhikara2025mem0,wang2023longmem,zhong2024memorybank,xu2025amem,anokhin2025arigraph,zhang2025associa} organizes and retrieves long-term memory; we study when a retrieved answer should be trusted. Selective prediction~\citep{elyaniv2010selective,geifman2017selective} uses pointwise signals (softmax, margin); the boundary cut is structural, using graph-neighbor agreement. The closest precedents are neighbor-agreement confidence methods: Trust Score~\citep{jiang2018trust} and deep-kNN confidence~\citep{papernot2018dknn} trust a prediction when its metric neighborhood, estimated from a labeled reference set, agrees with it. The boundary cut shares this intuition but uses a typed evidence graph and no labels; the semantic-kNN ablation (\S\ref{sec:graph-ablation}) shows that generic metric neighborhoods do not reproduce the typed-graph signal. Conformal prediction~\citep{vovk2005algorithmic,angelopoulos2023conformal,zhou2025conformalrandomobjects} provides distribution-free coverage via calibration; the boundary cut is training-free and could serve as a nonconformity score within a conformal pipeline. Laplacian regularization is classical in semi-supervised learning~\citep{belkin2006manifold,zhu2003semi}. Test-time compute work~\citep{snell2025testtime} shows abstention-guided selection outperforms extended search; we ask the analogous question for structured retrieval.

Spectral-summary work~\citep{martin2021predicting,martin2021implicit}
uses weight-matrix spectra as data-free quality proxies; our signal is
a Laplacian quadratic form over an evidence graph, with the
support-compatibility view of Section~\ref{sec:method} clarifying when
it should work.

\section{Limitations}
\label{sec:limitations}

\paragraph{Single positive benchmark family}
All positive aggregate results use MAB FactConsolidation. (Per-relation
lifts on CoDEx-S and MetaQA-3hop are exploratory; \S\ref{sec:per-relation-replication}.) We treat support compatibility as
a deployment condition, not as a universal property of structured
retrieval benchmarks.

\paragraph{Breadth of external validation}
The confirmatory cross-benchmark test (\S\ref{sec:cross-benchmark},
Appendix~\ref{app:e2}) spans only two independent external benchmark families,
LongMemEval and MetaQA; the LongMemEval temporal and semantic lanes are two
graph regimes within one family, not two independent families. The CoDEx-S
lane failed its reproduction gate (unseeded upstream ComplEx training) and is
exploratory only; the MAB single-hop and LoCoMo lanes were provenance-blocked.
Conclusions about cross-benchmark (non-)generalization should be read against
this limited external sample. The reported analyses use multiple preserved
evaluation populations, whose provenance is distinguished in
Table~\ref{tab:mab-provenance}. Some historical extraction and model states
are unavailable, limiting exact reconstruction of those runs.

\paragraph{Learned confidence}
The learned selector attains a higher point estimate, not statistically resolved in-distribution (GBT $0.814$ vs.\ boundary cut $0.785$ with full in-distribution labels; \S\ref{sec:cross-config-robustness}); portability across retrieval stacks is mixed (Table~\ref{tab:cross-config-bootstrap}). Percentile-rank features substantially mitigate but do not eliminate cross-stack degradation. Two points are specific
to this limitation. First, after purging repeated question texts from training, GBT has the higher point estimate on three of four held-out scales (Appendix~\ref{app:learned-confidence-extras}). Second, a post-hoc
complementarity check found no statistically resolved evidence that adding
the boundary cut to the GBT's five score-derived features improves out-of-fold AUPCC
($\Delta=+0.005$, 95\% CI grouped by exact question $[-0.005, +0.016]$); the two methods
differ in that the GBT requires correctness labels while the boundary cut does not.

\paragraph{Support-compatibility boundary}
Negative results on LongMemEval, LoCoMo, and CoDEx-S co-occurrence show that the boundary cut is not a generic graph confidence criterion: temporal adjacency, semantic kNN, and raw co-occurrence fail. The 50-example $R_\text{median}>2$ probe matches the observed usefulness labels 95\% of the time across eight families, including two prospectively-probed rejections with within-dataset heterogeneity (Appendices~\ref{app:gate}, \ref{app:per_relation}). The gate is a practical heuristic, not a validated predictor, and it requires labels. Label-free score computation should therefore be distinguished from validating whether a graph construction is suitable for deployment. Candidate-space cardinality, answer-space closedness, and typed-mask density are computable at deployment without labels and serve only as heuristic diagnostics (the closed-vs-open-vocabulary mechanism and the MetaQA candidate-filter/typed-mask substitution are detailed in \S\ref{sec:per-relation-replication}). This secondary diagnostic consumes correctness labels; its threshold was chosen on the original evaluated families, not independently validated as a general predictive criterion.

\paragraph{Other limitations and broader impact}
The mask depends on upstream entity extraction, and the perfect
zero-set precision is a clean-extraction property: under
entity-fragmentation noise on the 6k split (three seeds per rate),
zero-set precision degrades from $1.000$ to $0.907$, $0.874$, and
$0.715$ at 5\%, 10\%, and 20\% corrupted mentions, while retaining a
$+0.25$--$0.31$ precision lift over base accuracy at every rate.
Structural closure ($P=V$) is pipeline-dependent: it reflects the specific
extraction, candidate-neighborhood construction, and resolution procedure, and
may change under different such procedures. The boundary cut does not improve
base retrieval accuracy; the zero-cut subset covers only 17.75\% of
multi-hop queries; the search comparison is limited to multi-hop 6k (Table~\ref{tab:search}); and
the method applies to retrieved-item settings, not free-form text.
Because the cut and zero-cut decisions are deterministic functions
of the selected support and the graph weights (retrieval scores
enter upstream through path resolution), the boundary cut improves auditability
without additional model inference, but abstention criteria should not
be overtrusted.
Deployment also requires candidate-level retrieval scores, a graph
constructor that produces support-compatible edges, and a downstream
policy for abstained queries such as re-retrieval, escalation, or
clarification.

\section{Conclusion}
We study when typed evidence-graph structure predicts retrieval reliability and present a training-free abstention rule on selected-support boundaries. On MAB multi-hop, the rule identifies an empirically precise structural-closure regime produced by the upstream resolver. The decomposition separates closure, resolution state, topology, and edge weights: continuous weights add little on MAB, and the apparent raw-boundary signal outside closure largely reflects resolution-state composition. Cross-benchmark and relation-level analyses identify characteristics associated with useful abstention rather than a validated predictive criterion; the MAB normalization finding does not replicate externally. The status-matched score-baseline advantage is population-sensitive, while the regenerated population retains the closure regime and an advantage over the local QPP adaptations. GBT has a higher in-distribution point estimate with an unresolved paired difference, and transfer across retrieval stacks is mixed. These findings give an explicit abstention method and empirical guidance on its applicability and deployment tradeoffs without requiring correctness labels for the structural score. The central lesson is that the reliability of graph-based confidence depends on what the graph represents, not merely on its connectivity.

\bibliographystyle{ACM-Reference-Format}
\bibliography{references}

\FloatBarrier
\appendix

\section{Per-Split Results}
\label{app:splits}
\suppressfloats[t]
The tables and figure below summarize the per-split results.

\begin{table}[ht]
\centering
\caption{Per-split multi-hop AUPCC ($n{=}100$ per split).}
\label{tab:per-split-mh}
\begin{tabular}{lccccc}
\toprule
Split & Accuracy & Coherence & Smoothed & Raw & Margin \\
\midrule
6k   & 0.700 & \textbf{0.861} & 0.740 & 0.730 & 0.670 \\
32k  & 0.560 & \textbf{0.767} & 0.621 & 0.599 & 0.554 \\
64k  & 0.590 & \textbf{0.771} & 0.762 & 0.736 & 0.672 \\
262k & 0.420 & \textbf{0.661} & 0.495 & 0.468 & 0.355 \\
\midrule
All  & 0.568 & \textbf{0.785} & 0.677 & 0.653 & 0.547 \\
\bottomrule
\end{tabular}
\end{table}

\begin{table}[ht]
\centering
\caption{Per-split boundary-cut AUPCC by encoder (multi-hop, scorer held fixed at $(\mathbf{q}^\top\mathbf{v})^2$).}
\begin{tabular}{lccl}
\toprule
Split & Qwen $d{=}128$ & MiniLM $d{=}384$ & Winner \\
\midrule
6k   & \textbf{0.861} & 0.855 & Qwen (+0.006) \\
32k  & 0.767 & \textbf{0.769} & MiniLM (+0.002) \\
64k  & 0.771 & \textbf{0.776} & MiniLM (+0.005) \\
262k & 0.661 & \textbf{0.669} & MiniLM (+0.008) \\
\bottomrule
\end{tabular}
\end{table}

\begin{figure}[tbp]
\centering
\includegraphics[width=5in]{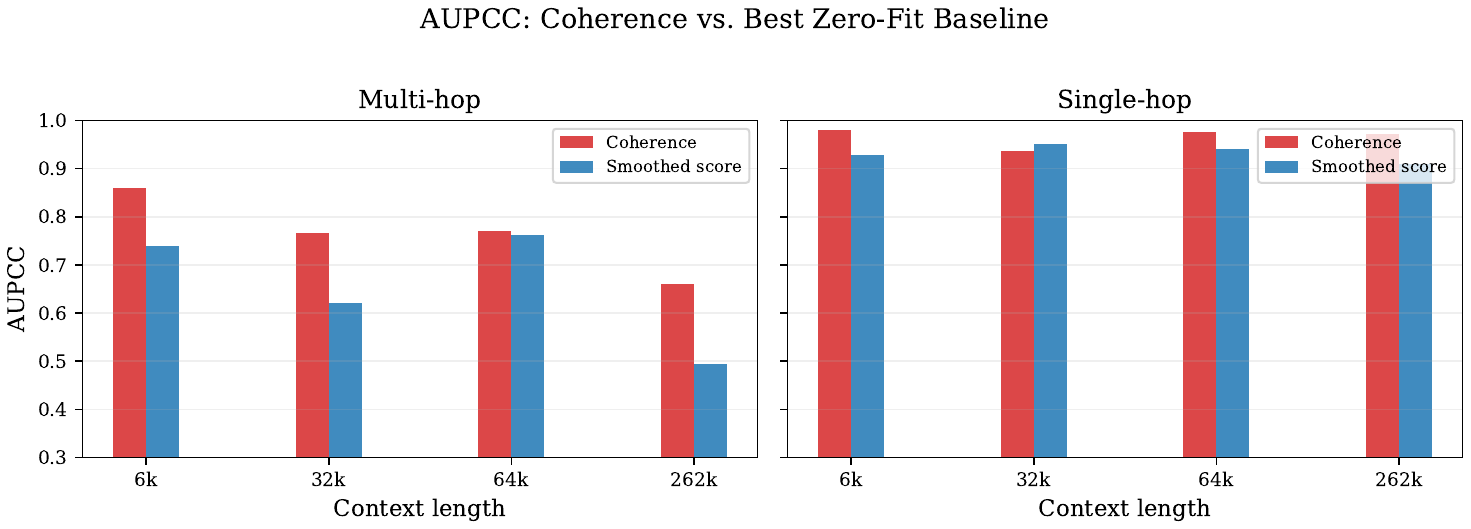}
\Description{Grouped bars compare the boundary cut and smoothed-score AUPCC at four context lengths. In multi-hop retrieval boundary-cut AUPCC is higher at all four lengths. In single-hop retrieval boundary-cut AUPCC is higher at 6k, 64k, and 262k, while smoothed score is higher at 32k (0.951 versus 0.936).}
\caption{AUPCC by context length. Boundary cut vs.\ the smoothed-score baseline across both regimes. The boundary cut has higher AUPCC at every multi-hop context length and at three of four single-hop context lengths; smoothed score is higher at single-hop 32k.}
\label{fig:aupcc-summary}
\end{figure}
\section{Scope of the Continuous Latent-Class Model}
\label{app:population-vs-selected}

This appendix develops the continuous latent-class model motivated in
\S\ref{sec:graph-structure} and clarifies its scope. The model is
qualitative intuition for support-compatible graph construction only, and it
is not instantiated in the reported experiments.

\paragraph{Latent-class decomposition}
The abstention properties are deterministic and apply to the weights
$w_{ij}$ of Eq.~\ref{eq:cut}, whether they are constructed directly or
derived from raw graph weights $s_{ij}$ through a support gate
$w_{ij}=s_{ij}m(q,o_i,o_j)$; the energy only sees the final weights. To
connect this to empirical behavior, consider a simple latent-class model.
Suppose each candidate belongs to an answer-support class $y_i$, and its
score is
\begin{equation}
    z_i = \mu_{y_i} + \varepsilon_i,
    \label{eq:latent-score}
\end{equation}
where $\mu_{y_i}$ is the class-level support and $\varepsilon_i$ is local
noise. Then
\begin{equation}
    E(\mathbf z)
    =
    \sum_{(i,j): y_i=y_j} w_{ij}(\varepsilon_i-\varepsilon_j)^2
    +
    \sum_{(i,j): y_i\neq y_j}
    w_{ij}(\mu_{y_i}-\mu_{y_j}+\varepsilon_i-\varepsilon_j)^2.
    \label{eq:support-decomp}
\end{equation}
The first term is score variation within the same support class. The second
term is score variation across different support classes. A
support-compatible graph constructor keeps little abstention weight in the
second term: it assigns large weights mostly to candidates whose support
scores should agree. A temporal graph, co-occurrence graph, or generic
similarity graph may instead assign large weights across support classes,
making the energy less informative for correctness.

\paragraph{Continuous-field concentration (intuition only)}
The latent-class model is stated for a continuous score field. On a connected
component, zero score-field energy forces all observed scores to be equal; if
that component contains candidates from two support classes separated by
$\Delta=|\mu_a-\mu_b|$, the noise must cancel the class-level gap. For
independent $\sigma$-sub-Gaussian noise,
\[
    \Pr(|\varepsilon_i-\varepsilon_j|\ge \Delta)
    \le 2\exp(-\Delta^2/(4\sigma^2)).
\]
We use this only as intuition for the generic continuous-field model. Real
retrieval scores are correlated through the shared encoder, scorer, and graph
construction, so the independence assumption does not hold.

\paragraph{Why the model does not explain the experiments}
Every reported boundary cut result verifies a binary selected-support indicator
(\S\ref{sec:setup}), so no reported energy is a continuous-field energy
generated under Eq.~\ref{eq:latent-score}: we do not use the reported
energies to estimate the model's parameters, and we do not use the model to
explain the MAB zero-cut precision. For the selected-support indicator the
relevant quantity is instead the boundary cut of Eq.~\ref{eq:cut}: what
matters is which edges cross from the selected support to its complement. The
latent-class partition is not, in general, the same partition as path
membership (an unselected candidate can share the answer-support class of a
selected one), so Eq.~\ref{eq:support-decomp} motivates graph construction
rather than explains the experimental zero set. The observed MAB mechanism is
instead deterministic and structural: in the regenerated population every
zero-cut case satisfies $P=V$, so the indicator field is constant and its
positive-weight boundary cut is empty (\S\ref{sec:zero-mechanism}), and
Remark~\ref{rem:boundary} requires no concentration argument.

We therefore retain the latent-class model only as motivation for
support-compatible graph construction (a constructor should avoid placing
weight on support-incompatible pairs), not as an account of the experimental
zero-cut subset. The paper does not claim any population-level probability
guarantee.

\section{Learned-Confidence Comparison: Extended Analysis}
\label{app:learned-confidence-extras}

\paragraph{Per-split bootstrap}
On the 262k split, boundary cut minus main-protocol GBT is $\Delta=+0.005$ (95\% exact-question bootstrap CI $[-0.073, +0.085]$), so the difference remains unresolved. The score ratio has mean impurity importance $0.691$ across the five corrected GBT folds. These results use the reconstructed mapping (Appendix~\ref{app:cross-config}).

\paragraph{AUROC}
On the same 400-query multi-hop set, AUROC is higher for GBT in aggregate (0.889 vs.\ 0.807), reflecting that the boundary cut is a precision-oriented selector strongest in the high-precision, low-coverage regime emphasized by AUPCC; AUROC weights all operating points equally.

\paragraph{Cross-scale generalization}
Training on the three shorter scales and testing on 262k, after purging 2 training rows whose question texts appear in the test set, yields GBT AUPCC $0.673$ versus the boundary cut's $0.661$. The corresponding within-distribution 262k GBT estimate is $0.656$. The 6k, 32k, and 64k held-out evaluations purge 21, 18, and 11 training rows, respectively. After these purges, GBT exceeds the boundary cut by point estimate on three of four held-out scales. All four paired intervals include zero. Details of the reconstructed mapping (Appendix~\ref{app:cross-config}) and training protocol are available from the author.

\begin{table}[htbp]
\centering\footnotesize
\caption{Question-disjoint leave-one-scale-out evaluation on headline MAB multi-hop ($N=100$ test rows per scale). Training rows sharing test-question text are purged. GBT-minus-boundary-cut intervals resample exact question texts under the reconstructed mapping (Appendix~\ref{app:cross-config}).}
\begin{tabular}{llllll}
\toprule
Test scale & Purged & GBT & Coherence & $\Delta$ & 95\% CI \\
\midrule
6k & 21 & 0.859 & 0.861 & -0.001 & [-0.077, +0.067] \\
32k & 18 & 0.794 & 0.767 & 0.027 & [-0.047, +0.095] \\
64k & 11 & 0.795 & 0.771 & 0.024 & [-0.057, +0.109] \\
262k & 2 & 0.673 & 0.661 & 0.012 & [-0.075, +0.094] \\
\bottomrule
\end{tabular}
\end{table}

\begin{table}[htbp]
\centering\footnotesize
\caption{Mean GBT impurity importance across five exact-question folds of headline MAB multi-hop ($N=400$), using the reconstructed mapping (Appendix~\ref{app:cross-config}). These are descriptive fold means, without bootstrap resampling.}
\begin{tabular}{ll}
\toprule
Feature & Importance \\
\midrule
Raw score & 0.058 \\
Smoothed score & 0.062 \\
Score margin & 0.139 \\
Second-highest score & 0.050 \\
Score ratio & 0.691 \\
\bottomrule
\end{tabular}
\end{table}

\section{Structural-Control Decomposition (E1)}
\label{app:e1}

This appendix reports the preregistered structural-control decomposition (E1)
summarized in \S\ref{sec:structural-controls}. The specification was frozen
before any comparative control outcome was computed.

\paragraph{Question} Does the boundary cut contain
selective-prediction information beyond structural closure $P=V$ and simple
graph-topology controls (boundary existence, boundary-edge count, selected
fraction, complement size)?

\paragraph{Population and provenance} The original graph-ablation cache no
longer exists on disk; the only per-query artifact retaining $V$, $P$, edges,
and weights is the regenerated zero-cut audit published in
\S\ref{sec:zero-mechanism} (pinned extraction model gpt-4.1-mini-2025-04-14;
$N=400$; $316$ resolved, $84$ unresolved chain-broken queries, all incorrect,
held at worst-confidence sentinels mirroring the boundary cut's convention; $70$
zero-cut, $67/71$ overlap with the original). This regenerated population
resolves $316$ of $400$ queries, whereas the original headline run resolves
$320$; the four-query difference is a property of the regenerated
extraction/resolution artifact (a different pinned extraction pass), not an
inconsistency in any headline number. The AUPCC evaluator is unchanged; the paired bootstrap now groups by verified exact question text. Reproduction gate: the
weighted cut reproduces AUPCC $0.7819$ on this population (vs.\ the paper's
$0.785$; $-0.0031$, the documented 70-vs-71 difference), with per-query parity
of recomputed vs.\ recorded energies to $5.7\times10^{-14}$.

\paragraph{Signals} $S_0$ weighted cut $E_w(P)=\sum_{i\in P,\,j\notin P} w_{ij}$;
$S_1$ closure indicator $\mathbf 1[P=V]$; $S_2$ boundary existence;
$S_3$ unweighted boundary count $E_1(P)$; $S_4$ selected fraction
$|P|/|V|$; $S_5$ complement size $|V\setminus P|$; $S_6$ boundary-weight
fraction (cut normalized by total neighborhood edge weight). Unresolved queries
receive, for every signal, a value strictly less confident than any resolved
value.

\paragraph{Populations and protocol} PA1: full $N=400$. PA2a (primary, literal
$P\neq V$): $N=330$, base accuracy $0.4697$. PA2b (secondary, resolved
$\wedge\ P\neq V$): $N=246$, base accuracy $0.6301$. The paired bootstrap
resamples question groups (cross-scale instances of a question together),
$10{,}000$ replicates, seed $42$;
$\Delta = \mathrm{AUPCC}(S_0) - \mathrm{AUPCC}(\text{control})$.

The structural-control estimates are consolidated in Table~\ref{tab:e1a-signals}.

\paragraph{Preregistered hypotheses}
\begin{itemize}
\item \textbf{H1 (SUPPORTED).} Full set: $\mathrm{AUPCC}(S_0)=0.7819$ vs.\
  $\mathrm{AUPCC}(S_1)=0.7199$; $\Delta=+0.0620$,\newline 95\% CI $[+0.030, +0.092]$,
  meeting the $\geq +0.02$ rule with the CI excluding $0$. The graded boundary
  carries information beyond the binary closure.
\item \textbf{H2 (SUPPORTED).} $P\neq V$ (PA2a): $\mathrm{AUPCC}(S_0)=0.5837$
  vs.\ base $0.4697$; margin $+0.114 \geq +0.05$. Boundary information survives
  removal of the closure cases.
\item \textbf{H3 (NOT SUPPORTED).} $P\neq V$ (PA2a):
  $\Delta(S_0-S_3)=-0.0064$, 95\% CI $[-0.020, +0.026]$; the point estimate is
  $\leq 0$, failing the preregistered rule. The weighted cut does not
  outperform the unweighted boundary count.
\end{itemize}
The overall verdict is partial support: the graded signal is not purely a
closure artifact (H1, H2), but the continuous edge magnitudes do not add value
over unweighted boundary topology (H3), and the selected fraction and
complement size reproduce the weighted cut's performance nearly everywhere. E1
is specific to MAB and does not speak to generalization; it also does not
establish the source of the residual $P\neq V$ separation, which the mixed
resolution states of that subset leave open.

\paragraph{Post-hoc observation (not confirmatory)} The boundary-weight
fraction $S_6$ (cut normalized by total neighborhood edge weight) numerically
exceeds every other signal on every population (full $0.8131$; $P\neq V$
$0.6575$) and has a higher point estimate than the raw weighted cut on PA2a, but the paired interval includes zero
($\Delta=-0.0738$, 95\% CI $[-0.146, +0.001]$). E1 preregistered no test
isolating whether this advantage comes from the weights or from the
normalization structure, so it is reported only as a hypothesis-generating
observation.

\section{Resolution-State and Normalization Diagnostic (E1A)}
\label{app:e1a}

This appendix reports the preregistered resolution-state and normalization
diagnostic (E1A) summarized in \S\ref{sec:resolution-state}. Its specification
was frozen (2026-08-13) before any E1A
comparative outcome was computed. E1A reuses the E1 per-query artifact and
evaluator unchanged; a parity gate reproduced the E1 values exactly (full
weighted cut $0.7819$; $P\neq V$ weighted $0.5837$, unweighted $0.5900$,
boundary-weight fraction $0.6575$). No LLM/API calls; no artifacts regenerated.

\paragraph{Question} How much of E1's non-closure separation survives
conditioning on resolution state, and does E1's post-hoc $S_6$ advantage come
from the continuous weights or from normalization itself (tested by the
unweighted analogue $S_8 = $ boundary-edge count $/$ total edge count)?

\paragraph{Populations} F (all, $N=400$); R (resolved, $N=316$); N (literal
$P\neq V$, $N=330$); \textbf{RN (primary: resolved $\wedge\ P\neq V$, $N=246$,
base accuracy $0.6301$)}; U (unresolved, $N=84$, descriptive only: all
incorrect, so any AUPCC on U is degenerate).

\paragraph{Unresolved-query audit} All $84$ unresolved queries lie in
$P\neq V$; $80$ carry a partial resolved prefix and $4$ have an empty path. Raw
cuts take the sentinel $10^9$ and the normalized fractions the sentinel $2.0$;
all $84$ are incorrect, so every structural-control signal ranks them last. On literal $P\neq V$ the
raw weighted cut sits $+0.114$ above base ($0.5837$ vs.\ $0.4697$), but on
resolved $P\neq V$ it sits \emph{below} base ($0.6000$ vs.\ $0.6301$): the
raw-cut residual in E1's literal population is attributable to the
unresolved-failure composition.

\begin{table}[htbp]
\centering\footnotesize
\caption{Structural-control AUPCC on regenerated MAB multi-hop: full ($N=400$), resolved ($N=316$), non-closure ($N=330$), and resolved non-closure ($N=246$). The second panel retains the additional full and non-closure controls. Paired intervals resample exact question texts.}
\label{tab:e1a-signals}
\begin{tabular}{lllllll}
\toprule
Population & $N$ & Accuracy & \shortstack{Weighted\\cut} & \shortstack{Unweighted\\cut} & \shortstack{Boundary-weight\\fraction} & \shortstack{Boundary-edge\\fraction} \\
\midrule
Full & 400 & 0.562 & 0.782 & 0.783 & 0.813 & 0.813 \\
Resolved & 316 & 0.712 & 0.822 & 0.819 & 0.861 & 0.861 \\
Non-closure & 330 & 0.470 & 0.584 & 0.590 & 0.657 & 0.662 \\
Resolved non-closure & 246 & 0.630 & 0.600 & 0.610 & 0.697 & 0.701 \\
\bottomrule
\end{tabular}
\par\medskip
\begin{tabular}{lcc}
\toprule
Additional control & Full ($N=400$) & Non-closure ($N=330$) \\
\midrule
Selected fraction & 0.783 & 0.579 \\
Complement size & 0.781 & 0.561 \\
Closure indicator $\mathbf 1[P=V]$ & 0.720 & 0.476 \\
\bottomrule
\end{tabular}
\end{table}

\paragraph{Signals and the 2$\times$2} $S_0$ weighted cut and $S_3$ unweighted
cut are the raw column; $S_6$ boundary-weight fraction and $S_8$ boundary-edge
fraction are the normalized column. $S_8$ is a post-E1 signal generated by E1's
post-hoc $S_6$ observation; its formula and all comparisons were frozen before
any E1A outcome, but any $S_8$ (A3/A4) finding is a within-dataset follow-up,
not external validation.

\paragraph{Paired bootstrap on RN}
All intervals below use $10{,}000$ replicates over question groups (seed $42$).
Each difference is the AUPCC of the first signal minus that of the second.
\begin{table}[htbp]
\centering\footnotesize
\caption{Paired AUPCC differences on regenerated MAB resolved non-closure queries (RN, $N=246$). Intervals use 10,000 paired bootstrap replicates grouped by exact question text.}
\begin{tabular}{lll}
\toprule
Comparison & $\Delta$ & 95\% CI \\
\midrule
Weighted cut - unweighted cut & -0.010 & [-0.027, +0.035] \\
Boundary-weight fraction - weighted cut & 0.097 & [-0.0002, +0.196] \\
Boundary-edge fraction - unweighted cut & 0.090 & [+0.007, +0.201] \\
Boundary-weight fraction - boundary-edge fraction & -0.004 & [-0.024, +0.016] \\
\bottomrule
\end{tabular}
\end{table}

\paragraph{Preregistered hypotheses (frozen labels and rules)}
\begin{itemize}
\item \textbf{A1 (residual raw signal survives the resolution filter): NOT
  SUPPORTED.} $\mathrm{AUPCC}(S_0\mid\text{RN})=0.6000 \leq$ base $0.6301$.
\item \textbf{A2 (raw weights still do not beat raw topology): SUPPORTED.}
  $\Delta(S_0-S_3)=-0.0104$, CI $[-0.027, +0.035]$ ($|\Delta|<0.02$, CI
  includes $0$).
\item \textbf{A3 (normalization improves the weighted signal): NOT SUPPORTED}
  (within-dataset follow-up). $\Delta(S_6-S_0)=+0.0968$, CI $[-0.0002, +0.196]$.
\item \textbf{A4 (weights vs.\ normalized topology): INDISTINGUISHABLE.}
  $\Delta(S_6-S_8)=-0.0037$, CI $[-0.024, +0.016]$, $|\Delta|<0.02$.
\end{itemize}
Confirmatory analyses are A1 and A2 (raw signals, E1-defined); A3 and A4 are
frozen but within-dataset follow-ups involving the post-E1 signal $S_8$. The
combined E1$+$E1A reading: continuous edge weights are unnecessary (raw and
normalized); the strict $100\%$-precision regime is structural closure $P=V$; the
non-closure residual is largely resolution-state composition; and among resolved
non-closure queries a normalized boundary fraction (relative boundary
exposure) carries a moderate signal ($\sim+0.07$ over base) that is an
MAB-specific hypothesis requiring external validation. Caveats: single dataset,
$N=246$, wide CIs, and $S_6/S_8$ evaluated on their motivating data.

\paragraph{Provenance} E1A operates on the same audited
per-query artifact as E1 (verified by an identical integrity hash), with no
re-extraction or recomputation.

\section{Cross-Benchmark Validation (E2)}
\label{app:e2}

This appendix reports the preregistered cross-benchmark validation (E2)
summarized in \S\ref{sec:cross-benchmark}. Its specification was frozen
(2026-08-13) after the provenance/reproduction
gates but before any comparative $S_8/S_6/S_3/S_0$ outcome was computed. No
LLM/API/network calls were made (\texttt{HF\_HUB\_OFFLINE}); no E1/E1A artifacts
were modified.

\paragraph{Question and frozen hypotheses} Does the MAB-generated hypothesis that normalized unweighted boundary topology
$S_8$ (boundary-edge count $/$ total edge count) is a transferable zero-fit
confidence signal generalize outside MAB? Frozen hypotheses: \textbf{E2-H1} (normalization generalizes: $S_8$ beats
raw $S_3$ across external families); \textbf{E2-H2} ($S_8$ exceeds base accuracy
on $\geq 1$ confirmatory external population); \textbf{E2-H3} (weights not
required in the abstention statistic: no confirmatory instantiation shows
$\Delta(S_6-S_8)\geq 0.02$ with CI excluding $0$); \textbf{E2-H4}
(methodological: report full vs.\ resolved wherever a failure state exists).

\paragraph{Reproduction-gate protocol} Each lane recomputes the benchmark's own
published boundary cut result from its existing pipeline (selector, graph, $k$,
correctness) through the paper's evaluator; a lane is confirmatory only if this
reproduces the published value. Weight conclusions concern the post-selection
abstention statistic only; every external pipeline uses graph-smoothed
selection upstream.

\begin{table}[t]
\centering
\small
\caption{E2 reproduction gates. Confirmatory lanes reproduce the published
result; CoDEx-S fails (unseeded upstream training); two lanes are
provenance-blocked.}
\label{tab:e2-gates}
\begin{tabular}{llll}
\toprule
Benchmark / graph & $N$ & Reproduced vs published & Gate \\
\midrule
LongMemEval / temporal & 500 & $0.736$ vs $0.736$ & PASS \\
LongMemEval / semantic & 500 & $0.682$ vs $0.682$ & PASS \\
MetaQA-3hop / typed & 250 & $\Delta{=}{+}0.020$ vs ${+}0.023$ & PASS \\
CoDEx-S / co-occurrence & 1828 & $0.237$ vs $0.373$ & FAIL (exploratory) \\
MAB single-hop & --- & --- & BLOCKED (provenance) \\
LoCoMo & --- & --- & BLOCKED (provenance) \\
\bottomrule
\end{tabular}
\end{table}

\paragraph{Confirmatory signal AUPCC} On each gate-passing lane's full
population (base accuracy in parentheses): raw weighted $S_0$, raw unweighted
$S_3$, normalized weighted $S_6$, normalized unweighted $S_8$.

\begin{table}[htbp]
\centering\footnotesize
\caption{Confirmatory E2 AUPCC on external populations: LongMemEval temporal and semantic ($N=500$ each) and MetaQA-3hop ($N=250$). Paired intervals resample each benchmark\textquotesingle s external query identifier. These populations are separate from headline and regenerated MAB.}
\label{tab:e2-signals}
\begin{tabular}{lllll}
\toprule
Population & \shortstack{Weighted\\cut} & \shortstack{Unweighted\\cut} & \shortstack{Boundary-weight\\fraction} & \shortstack{Boundary-edge\\fraction} \\
\midrule
LongMemEval temporal & 0.736 & 0.745 & 0.733 & 0.736 \\
LongMemEval semantic & 0.682 & 0.779 & 0.853 & 0.774 \\
MetaQA-3hop & 0.324 & 0.324 & 0.283 & 0.283 \\
\bottomrule
\end{tabular}
\end{table}

\paragraph{Paired bootstrap per lane}
The following differences use $10{,}000$ replicates (seed $42$) and 95\%
percentile confidence intervals.
\begin{table}[htbp]
\centering\footnotesize
\caption{External E2 paired AUPCC differences: LongMemEval temporal and semantic ($N=500$ each), MetaQA-3hop ($N=250$), and exploratory CoDEx-S ($N=1828$). Intervals use 10,000 replicates grouped by external query identifier. CoDEx-S failed its reproduction gate.}
\begin{tabular}{llll}
\toprule
Lane & Comparison & $\Delta$ & 95\% CI \\
\midrule
LongMemEval temporal & boundary-edge fraction $-$ unweighted cut & -0.009 & [-0.026, +0.015] \\
LongMemEval temporal & boundary-edge fraction $-$ weighted cut & 0.001 & [-0.043, +0.044] \\
LongMemEval temporal & boundary-weight fraction $-$ boundary-edge fraction & -0.003 & [-0.045, +0.039] \\
LongMemEval semantic & boundary-edge fraction $-$ unweighted cut & -0.005 & [-0.017, +0.007] \\
LongMemEval semantic & boundary-edge fraction $-$ weighted cut & 0.093 & [+0.041, +0.146] \\
LongMemEval semantic & boundary-weight fraction $-$ boundary-edge fraction & 0.078 & [+0.048, +0.111] \\
MetaQA-3hop & boundary-edge fraction $-$ unweighted cut & -0.041 & [-0.095, +0.022] \\
MetaQA-3hop & boundary-edge fraction $-$ weighted cut & -0.041 & [-0.095, +0.022] \\
MetaQA-3hop & boundary-weight fraction $-$ boundary-edge fraction & 0.000 & [+0.000, +0.000] \\
CoDEx-S & boundary-edge fraction $-$ unweighted cut & -0.001 & [-0.009, +0.007] \\
CoDEx-S & boundary-edge fraction $-$ weighted cut & -0.003 & [-0.010, +0.006] \\
CoDEx-S & boundary-weight fraction $-$ boundary-edge fraction & 0.003 & [-0.003, +0.009] \\
\bottomrule
\end{tabular}
\par\smallskip\begin{minipage}{0.95\linewidth}\scriptsize CoDEx-S remains exploratory because its reproduction gate failed.\end{minipage}
\end{table}

\paragraph{Frozen hypothesis verdicts}
\begin{itemize}
\item \textbf{E2-H1 (normalization generalizes): NOT SUPPORTED.} Both
  confirmatory external families have $\Delta(S_8-S_3)\leq 0$ by point estimate
  (temporal $-0.0088$, semantic $-0.0049$, MetaQA $-0.0407$); none positive.
\item \textbf{E2-H2 ($S_8$ above base somewhere external): SUPPORTED (weakly).}
  On LME semantic, $S_8=0.7744$, above base accuracy $0.748$, but the raw count $S_3$ ($0.7793$)
  and $S_6$ ($0.8528$) are both higher, so $S_8$ is above base without being the
  operative signal; elsewhere $S_8$ is at or below base.
\item \textbf{E2-H3 (weights not required in the abstention statistic): NOT
  SUPPORTED.} On the semantic-graph instantiation $S_6$ beats $S_8$ by $+0.0785$,
  CI $[+0.048, +0.111]$ (material, CI excludes $0$).
\item \textbf{E2-H4 (failure-state separation): NOT APPLICABLE IN SAMPLE.} No
  evaluated benchmark produced an unresolved/failure state (MetaQA: $0/250$;
  others resolve by construction), so full $\equiv$ resolved everywhere.
\end{itemize}
\textbf{Frozen cross-benchmark verdict: MAB-specific.} The MAB normalization hypothesis did not generalize across benchmarks;
behavior is regime-dependent; one external semantic regime shows
that continuous weights can matter.

\paragraph{External-sample caveat} Confirmatory evidence spans two independent
external families (LongMemEval, MetaQA); LongMemEval temporal and semantic are
two graph regimes within one family, not two independent families.

\paragraph{Exploratory (gate-failed) CoDEx-S} Reported for completeness only;
the reproduction gate failed ($0.2373$ vs published $0.3727$; unseeded ComplEx
initialization), so no confirmatory conclusion is licensed. Full-population
AUPCC: base $0.3255$, $S_0$ $0.2373$, $S_3$ $0.2355$, $S_6$ $0.2370$, $S_8$
$0.2344$; $\Delta(S_8-S_3)=-0.0011$ $[-0.009, +0.007]$. Mild positive $S_8$
deltas appear on a few closed/dense relations, directionally reminiscent of the
closed-vocabulary pattern, but this run does not reproduce the published
pipeline and licenses no conclusion.

\paragraph{Provenance-blocked lanes} MAB single-hop: the pinned LLM-extraction
cache covers multi-hop splits only, so per-query graphs are unavailable and
regeneration would require new LLM calls (prohibited). LoCoMo: the session-content
file is absent from disk (only questions present) and re-download is prohibited.
Neither lane is interpreted as E2 evidence.

\paragraph{Confirmatory vs.\ post-hoc} Confirmatory: the frozen E2-H1--H4
adjudications above. Post-hoc / hypothesis-generating: the mechanistic reading
of the LME-semantic $S_6 > S_8$ result (that weight semantics may matter when
edge weights encode task-relevant compatibility), which is a single-instantiation
observation in the opposite direction of the E1A weights finding and cannot
rescue the preregistered hypothesis.

\paragraph{Provenance} The evaluator is unchanged; E2 bootstrap clusters use the external query identifier directly, with 10{,}000 replicates and seed 42.

\section{Linear-Scorer Ablation}
\label{app:raw-smoothed}

We ablate the default squared scorer $(\mathbf{q}^\top\mathbf{v})^2$ by recomputing all selectors on the linear scorer $\mathbf{q}^\top\mathbf{v}$. Table~\ref{tab:raw-smoothed-app} reports boundary cut and three score-only baselines on this linear-scorer field. The boundary cut remains the strongest zero-fit selector. This is a scorer-function ablation, not a smoothing ablation: $E(\mathbf z)=\mathbf z^\top L\mathbf z$ depends only on the graph and the node field; $\gamma$ enters at most upstream through path resolution, and Appendix~\ref{app:gamma} shows the signal is unchanged across $\gamma$.

\begin{table}[ht]
\centering
\caption{Selector comparison on multi-hop MAB under the linear scorer $\mathbf{q}^\top\mathbf{v}$ (alternative to the default $(\mathbf{q}^\top\mathbf{v})^2$).}
\label{tab:raw-smoothed-app}
\begin{tabular}{lcc}
\toprule
Signal & Aggregate AUPCC & 262k point estimate \\
\midrule
Coherence & \textbf{0.775} & \textbf{0.624} \\
Smoothed score & 0.688 & 0.518 \\
Raw score & 0.667 & 0.490 \\
Margin & 0.612 & 0.435 \\
\bottomrule
\end{tabular}
\end{table}

\section{Typed-Edge Robustness}
\label{app:robustness}

Table~\ref{tab:robustness-app} reports how the boundary cut changes when the typed graph is deliberately corrupted. Missing edges hurt more than added spurious edges, but performance degrades gracefully rather than catastrophically. The uncorrupted baseline AUPCC ($0.7495$) reported here comes from this perturbation experiment's own seed/protocol and differs by ${\sim}0.016$ from the typed-graph entry in Table~\ref{tab:graph-ablation-app} ($0.7652$); deltas within Table~\ref{tab:robustness-app} are internally consistent against this baseline.

\begin{table}[ht]
\centering
\caption{Robustness of the boundary cut to graph corruption on cached multi-hop MAB neighborhoods.}
\label{tab:robustness-app}
\begin{tabular}{lcc}
\toprule
Perturbation & AUPCC & Delta \\
\midrule
Baseline & 0.750 & --- \\
Drop 10\% edges & 0.716 $\pm$ 0.005 & $-0.034$ \\
Drop 20\% edges & 0.717 $\pm$ 0.005 & $-0.033$ \\
Drop 30\% edges & 0.720 $\pm$ 0.002 & $-0.029$ \\
Add 10\% spurious & 0.743 $\pm$ 0.001 & $-0.007$ \\
Add 20\% spurious & 0.743 $\pm$ 0.001 & $-0.006$ \\
\bottomrule
\end{tabular}
\end{table}

\section{Gamma Invariance}
\label{app:gamma}

The boundary cut $E(\mathbf z) = \mathbf z^\top L \mathbf z$ depends on the graph Laplacian and the node field; the smoothing hyperparameter $\gamma$ does not enter the cut directly and can affect it at most upstream through path resolution, so the abstention signal is expected to be stable in $\gamma$. Table~\ref{tab:gamma-sweep} confirms this empirically. The smoothed-score baseline shifts with $\gamma$, but the boundary cut AUPCC and the zero-cut subset are nearly unchanged.

\begin{table}[ht]
\centering
\caption{Gamma sweep on multi-hop MAB (400 queries). Boundary-cut AUPCC is numerically stable across the tested $\gamma$ values (0.765--0.766), with the same 71 zero-energy queries at 100\% precision throughout.}
\label{tab:gamma-sweep}
\begin{tabular}{rccc}
\toprule
$\gamma$ & Coherence AUPCC & ZE queries & ZE precision \\
\midrule
0.000 & 0.766 & 71 & 1.000 \\
0.050 & 0.765 & 71 & 1.000 \\
0.100 & 0.765 & 71 & 1.000 \\
0.200 & 0.765 & 71 & 1.000 \\
\bottomrule
\end{tabular}
\end{table}

\section{AUROC Comparison}
\label{app:auroc}

Table~\ref{tab:auroc-full} reports AUROC for all selectors, complementing the AUPCC values in Table~\ref{tab:learned}. AUROC measures overall discrimination between correct and incorrect queries without the coverage-weighting of AUPCC. GBT achieves the highest AUROC ($0.889$), followed by the boundary cut ($0.807$). The boundary cut's AUROC advantage over score baselines is consistent with its AUPCC advantage: the graph signal provides correctness information beyond what scores alone capture.

\begin{table}[htbp]
\centering\footnotesize
\caption{AUROC for headline MAB multi-hop ($N=400$ overall; $N=100$ per context scale). GBT and LR use folds grouped by exact question under the reconstructed mapping (Appendix~\ref{app:cross-config}); paired uncertainty uses exact-question resampling.}
\label{tab:auroc-full}
\begin{tabular}{llllll}
\toprule
Signal & All & 6k & 32k & 64k & 262k \\
\midrule
GBT & 0.889 & 0.912 & 0.888 & 0.885 & 0.850 \\
Coherence & 0.807 & 0.753 & 0.793 & 0.792 & 0.845 \\
Smoothed score & 0.645 & 0.586 & 0.605 & 0.732 & 0.617 \\
Raw score & 0.611 & 0.573 & 0.584 & 0.683 & 0.578 \\
LR & 0.620 & 0.586 & 0.592 & 0.655 & 0.605 \\
Score margin & 0.413 & 0.458 & 0.447 & 0.508 & 0.351 \\
\bottomrule
\end{tabular}
\end{table}

\section[R-median Gate: A Labeled Diagnostic]{$R_\text{median}$ Gate: A Small Labeled Diagnostic}
\label{app:gate}

This gate is a small labeled diagnostic (an audit for whether the boundary cut is likely to help on a new graph construction), not part of a label-free deployment mechanism: it consumes ground-truth labels in the probe set. To test whether support compatibility can be assessed prospectively from such a probe, we repeatedly sample tiny labeled probe sets (20, 30, and 50 examples) from each of eight graph families and apply a simple decision rule: predict ``boundary cut likely useful'' if $R_\text{median} > 2$, where $R_\text{median}$ is the ratio of median cut on incorrect vs.\ correct answers within the probe. We compare this prediction against the full-run ground truth (Table~\ref{tab:gate}). This secondary diagnostic consumes correctness labels; its threshold was chosen on the original evaluated families, not independently validated as a general predictive criterion.

\begin{table}[ht]
\centering
\caption{Prospective $R_\text{median}$ gate (1{,}000 trials per cell, threshold $=2$). Decision accuracy is the fraction of trials where the gate matches the full-run ground truth. The gate is conservative: false-positive rate $<1\%$ at probe size 50, meaning it rarely approves a graph that should be rejected. This secondary diagnostic consumes correctness labels; its threshold was chosen on the original evaluated families, not independently validated as a general predictive criterion.}
\label{tab:gate}
\begin{tabular}{lcccccc}
\toprule
 & \multicolumn{2}{c}{Probe 20} & \multicolumn{2}{c}{Probe 30} & \multicolumn{2}{c}{Probe 50} \\
\cmidrule(lr){2-3}\cmidrule(lr){4-5}\cmidrule(lr){6-7}
Family (truth) & Acc & Med $R_m$ & Acc & Med $R_m$ & Acc & Med $R_m$ \\
\midrule
MAB typed (useful)        & 0.860 & 8.140 & 0.920 & 8.580 & 0.950 & 7.530 \\
MAB sem.\ kNN (useful)    & 0.550 & 2.250 & 0.610 & 2.390 & 0.680 & 2.500 \\
LongMemEval temp.\ (not)  & 0.970 & 1.000 & 0.980 & 0.970 & 1.000 & 0.990 \\
LongMemEval kNN (not)     & 0.850 & 0.650 & 0.920 & 0.650 & 0.970 & 0.590 \\
LoCoMo temp.\ (not)       & 1.000 & 1.020 & 1.000 & 1.030 & 1.000 & 1.030 \\
CoDEx-S co-occ.\ (not)    & 0.990 & 0.580 & 1.000 & 0.590 & 1.000 & 0.630 \\
CoDEx-S typed (not)       & 0.970 & 1.010 & 0.980 & 0.990 & 1.000 & 1.010 \\
MetaQA-3hop typed (not)   & 0.920 & 1.030 & 0.970 & 1.040 & 1.000 & 1.000 \\
\midrule
Overall                    & 0.890 &      & 0.920 &      & 0.950 &      \\
\bottomrule
\end{tabular}
\end{table}

At probe size 50, the gate achieves 95\% overall accuracy across eight families with a false-positive rate of 0.5\%. The only substantial source of error is the borderline case (MAB semantic kNN, full-run $R_\text{median}{=}2.75$, just above the threshold). The four clearly misaligned graphs in the original probe set are rejected at ${\geq}97\%$ accuracy even at probe size 20. This secondary diagnostic consumes correctness labels; its threshold was chosen on the original evaluated families, not independently validated as a general predictive criterion.

This is a heuristic, not a validated theorem. The threshold of 2 and the probe sizes were chosen post hoc on the original six families used for evaluation. Nevertheless, the result suggests that a small labeled probe can provide a useful go/no-go signal for deployment of the abstention rule.

We add CoDEx-S with a typed entity--relation mask and MetaQA-3hop as the seventh and eighth prospectively probed families. Both are correctly rejected: CoDEx-S typed has mean $R_{\mathrm{median}} = 1.01$ (std 0.23 across five probe-50 seeds), MetaQA-3hop has mean $R_{\mathrm{median}} = 1.04$, and decision accuracy is ${\geq}92\%$ at every probe size. This extends the diagnostic to two previously unprobed families: the threshold of 2 was chosen on the original six evaluated families. It does not establish a general predictive criterion. The per-relation breakdown of both rejected datasets (Appendix~\ref{app:per_relation}) shows that aggregate rejection masks within-dataset heterogeneity. CoDEx-S contains a relation that would individually pass the threshold-2 gate (P463 ``member of'' with $R_{\mathrm{median}}=52$); MetaQA-3hop is uniformly subthreshold (max $R_{\mathrm{median}}=1.46$ for \texttt{in\_language}) yet still contains relations with positive boundary-cut lifts. This secondary diagnostic consumes correctness labels; its threshold was chosen on the original evaluated families, not independently validated as a general predictive criterion.

\section{Support Compatibility as a Per-Relation Property}
\label{app:per_relation}

Both CoDEx-S typed and MetaQA-3hop are correctly rejected by the $R_{\mathrm{median}}$ gate at the aggregate level (Appendix~\ref{app:gate}), yet both contain relations on which the abstention rule works. This appendix reports per-answer-relation breakdowns for each dataset and identifies the structural conditions under which the cut signal is informative. This secondary diagnostic consumes correctness labels; its threshold was chosen on the original evaluated families, not independently validated as a general predictive criterion.

\subsection{CoDEx-S Typed}

The aggregate CoDEx-S result (boundary cut AUPCC 0.3727 vs.\ raw score 0.5430) masks substantial heterogeneity across relations. Table~\ref{tab:codex_per_relation} reports boundary cut performance for the seven most-populated relations in the CoDEx-S typed evaluation.

For CoDEx-S typed, compatible pairs are constructed from relation ids and
hand-specified compositional relation families. The mask uses relation
metadata only; it is not learned from correctness labels and does not
read test-time answer labels.

Three patterns emerge:

\paragraph{Support-compatible relations}
P463 (``member of'') achieves AUPCC 0.700 with $R_{\mathrm{median}} = 52$ and a zero-cut subset of 89 queries at 76.4\% precision. This relation behaves like the MAB typed graph: organizational membership creates dense, support-compatible neighborhoods where the boundary cut separates correct from incorrect answers. This secondary diagnostic consumes correctness labels; its threshold was chosen on the original evaluated families, not independently validated as a general predictive criterion.

\paragraph{Sparse-mask collapse}
P530 (``diplomatic relation'') assigns zero cut to all 287 queries because the typed mask produces disconnected local subgraphs: every neighborhood is a single node or pair with no typed edges. Zero-cut precision equals base accuracy (0.244), confirming that the cut signal is trivially uninformative. This is a graph-construction failure, not a boundary-cut failure: the mask is too restrictive for this relation's neighborhood structure.

\paragraph{Dense but misaligned}
P106 (``occupation,'' 593 queries) has dense typed neighborhoods from the Person-Role family, but the boundary cut (0.215) actively underperforms raw score (0.313) with $R_{\mathrm{median}} = 0.88 < 1$. The typed edges connect candidates whose support \emph{should not} agree under occupation queries: a person's employer and field of work do not reliably predict whether a specific occupation candidate is correct. This secondary diagnostic consumes correctness labels; its threshold was chosen on the original evaluated families, not independently validated as a general predictive criterion.

These patterns refine the support-compatibility thesis from a binary dataset-level property to a continuous, relation-level property. A dataset can contain both support-compatible and support-incompatible relations; the aggregate result reflects the mixture. Among the seven relations above, only P463 exceeds the gate threshold of 2, and it is the only relation where the boundary cut outperforms raw score. These CoDEx-S per-relation values are \emph{artifact-conditional}: the exact historical ComplEx initialization is not reproducibly recoverable and the original per-example predictions were not persisted, so per-relation bootstrap intervals could not be reconstructed; the point estimates are reported as illustrative exploratory evidence only.

\subsection{MetaQA-3hop}

MetaQA-3hop \citep{zhang2018variational} provides an independent test of the per-relation pattern: a different domain (movie KG with 9 relations), a different scoring channel (MiniLM bi-encoder, $d{=}384$), and a different retrieval setup (per-question subgraph instead of trained KG embedding scorer). We sample 250 queries stratified by extracted question--relation set and apply the same boundary-cut pipeline with an answer-type candidate filter (entities appearing as objects of the inferred answer relation in the subgraph are scored).

The aggregate result mirrors CoDEx-S typed: the boundary cut narrowly improves over raw score by $+0.023$ AUPCC (95\% bootstrap CI $[-0.068, +0.104]$, 2{,}000 iterations), $R_{\mathrm{median}} = 1.10$, and the gate at threshold 2 correctly rejects deployment in aggregate. Per-relation, behavior is heterogeneous, though the per-relation intervals are wide and only \texttt{in\_language} excludes zero (Table~\ref{tab:metaqa_per_relation}). This secondary diagnostic consumes correctness labels; its threshold was chosen on the original evaluated families, not independently validated as a general predictive criterion.

For \texttt{in\_language}, the boundary-cut lift is $+0.230$ AUPCC at base accuracy 0.435. Language candidates per query have median 4, compared with medians 15--33 for person-name candidates.

\paragraph{Candidate-type filtering and edge-type masking are substitutes}
The MetaQA pipeline applies an answer-type candidate filter upstream: only entities appearing as objects of the inferred answer relation in the subgraph are scored. Typed and untyped graphs produce identical top-1 picks on 243 of 250 queries; the aggregate AUPCC delta is $-0.003$. 

\section{Question Identity and Cross-Configuration Evaluation}
\label{app:cross-config}

The earliest transfer harness assigned a separate group to each pair
(context scale, positional query index). A later correction grouped rows by the
positional index alone, assuming that the same index referred to the same question
at all four scales. Inspection of the actual question lists contradicts that
assumption. The reconstructed headline mapping contains 374 distinct exact question
texts across 400 evaluations. There are 26 repeated texts, each appearing twice,
including repetitions at different indices. Neither the composite row identifier
nor the positional index is the appropriate identity key for those repetitions.
The previous claim that index grouping had corrected question leakage was therefore
unsupported; its reported 99/100 leakage count measured index reuse, not verified
question reuse.

The repaired evaluation uses five-fold cross-validation grouped by exact question
text, keeping every occurrence together across scales and configurations. Existing
features, labels and estimator hyperparameters are retained. The transfer harness
uses numeric sorting by (context\_length,query\_index) and no feature standardization.
This ordering was verified by reproducing all 18 archived raw/population-rank cells
under their original index grouping. The main learned-confidence experiment instead
uses native CSV order and training-fold StandardScaler. The separate native-order
unscaled sensitivity (historically 0.8200, repaired to 0.812) is retained as a protocol sensitivity, not a second primary
estimate. The main comparison uses its own repaired paired interval; the separate unscaled evaluation is retained only as a protocol sensitivity (details available from the author). Row order and scaling remain protocol
differences, but neither justifies calling the earlier folds question-disjoint.

The mapping from historical feature rows to question text uses the reconstructed mapping described in this appendix, obtained by
(context\_length,query\_index) from the available dataset snapshot. Its exact data
content and question-list hashes are recorded, and agree with the verified
regenerated bundle. The original headline run did not record a dataset revision,
so this does not establish independently authenticated historical identity.

Every repaired paired interval uses 10000 exact-question bootstrap replicates with
seed 42, conditional on the fixed out-of-fold predictions. The archived transfer
correction report specifies 2000 replicates, while the earlier manuscript described 10000; the
repair resolves this discrepancy prospectively and does not relabel the old run.

The archived rank-normalization diagnostic ranked each feature column across all 400
queries within each configuration, before cross-validation, with positional tie
breaks. This is a transductive population-rank transformation, not a within-query
rank. Its repaired result is reported separately. Literal within-query candidate
percentiles cannot be computed from the headline scalar-feature artifacts, which
lack candidate-score vectors. Ranking five heterogeneous scalar features against
one another would be a different, unsupported definition. The 8-of-9 population-rank
finding must not be presented as evidence for within-query normalization.

\paragraph{Transfer data}
The repaired AUROC matrix and the unchanged feature-scale distributions accompany the grouping history.
\begin{table}[htbp]
\centering\footnotesize
\caption{GBT transfer AUROC on headline MAB multi-hop ($N=400$ per target). Rows identify training configurations and columns deployment configurations. Folds and paired resampling use exact question texts under the reconstructed mapping (Appendix~\ref{app:cross-config}).}
\label{tab:cross-config-auroc}
\begin{tabular}{llll}
\toprule
Train / test & \textsc{base} & \textsc{minilm} & \textsc{linear} \\
\midrule
\textsc{base} & 0.891 & 0.841 & 0.857 \\
\textsc{minilm} & 0.761 & 0.909 & 0.753 \\
\textsc{linear} & 0.798 & 0.744 & 0.868 \\
\bottomrule
\end{tabular}
\end{table}

\begin{table}[ht]
\centering
\caption{\texttt{raw\_score} distribution per configuration ($n{=}400$ queries).}
\label{tab:cross-config-feat-scale}
\small
\begin{tabular}{lcccc}
\toprule
Configuration & Mean & Std & Range & Median \\
\midrule
\textsc{base}    & 0.008 & 0.002 & $[0.000, 0.009]$ & 0.008 \\
\textsc{minilm}  & 0.003 & 0.001 & $[0.000, 0.003]$ & 0.003 \\
\textsc{linear}  & 0.390 & 0.131 & $[0.000, 0.625]$ & 0.406 \\
\bottomrule
\end{tabular}
\end{table}

\end{document}